\documentclass[oldversion]{aa}

\usepackage{natbib}
\usepackage[french,english]{babel}

\usepackage[latin1]{inputenc}
\usepackage[T1]{fontenc}
\usepackage{xspace,verbatim}
\usepackage{graphicx,rotating,array,subfigure,color}
\usepackage{txfonts}
\usepackage[bookmarks=true,bookmarksopen=true]{hyperref}

\newif\ifendfloat
\endfloatfalse    

\providecommand{\unit}[1]{\ensuremath{\:\mathrm{#1}}}
\providecommand{\textind}[1]{{\ensuremath{\textrm{\scriptsize{#1}}}}}

\def\eg{{\it e.g.}\ }

\def\ie{{\it i.e.}\ }

\newcommand{\ud}{\ensuremath\,\mathrm{d}}

\def\lyeps{Ly\,$\varepsilon$\xspace}
\def\svi{S\,\textsc{vi}\xspace}
\def\svii{S\,\textsc{vi}~93.3\,nm\xspace}
\def\svij{S\,\textsc{vi}~94.4\,nm\xspace}
\def\sumer{SoHO/SUMER\xspace}

\newcommand\moment[1]{(#1)}

\begin{document}
\selectlanguage{english}

\title{A statistical study of SUMER spectral images: \\
  events, turbulence, and intermittency}
\author{E. Buchlin\inst{1,2}
  \and J.-C. Vial\inst{1}
  \and P. Lemaire\inst{1}
}

\titlerunning{A statistical study of SUMER spectral images}

\offprints{E. Buchlin, \protect\url{eric@arcetri.astro.it}}

\institute{
  Institut d'Astrophysique Spatiale, CNRS -- Universit\'e Paris-Sud,
  B\^at.~121, 91405 Orsay Cedex, France
  \and
  Dipartimento di Astronomia e Scienza dello Spazio, Universit\`a
  di Firenze, Largo E. Fermi 2, 50125 Firenze, Italy
}

\date{Received\,:  / Revised date\,:}

\abstract{ We analyze a series of full-Sun observations, which was performed
  with the \sumer instrument between March and October 1996. Some parameters
  (radiance, shift and width) of the \svii, \svij, and \lyeps line profiles
  were computed on board.  Radiances and line-of-sight velocities in a large
  central region of the Sun are studied statistically: distributions of
  solar structures, field Fourier spectra and structure functions are
  obtained. The structures have distributions with power-law tails, the
  Fourier spectra of the radiance fields also display power laws, and the
  normalized structure functions of the radiance and velocity fields
  increase at small scales. These results support the idea of the existence
  of small scales, created by turbulence, and of intermittency of the
  observed fields.  These properties may provide insight into the processes
  needed for heating the transition region, or, if confirmed in the corona,
  the corona itself. The difficulties encountered in this analysis,
  especially for the velocity data, underline the needs for sensitive
  ultraviolet imaging spectrometers.

  \keywords{
    Sun: Transition Region, flares; Turbulence
  }

}

\maketitle

\section{Introduction}

The mechanism heating the solar corona to millions of degrees kelvin remains
an open problem, but it is generally understood that a great part of the
energy dissipation must occur at scales that are smaller than the structures
that can be resolved by observations ($\approx 100\unit{km}$), perhaps as
small as $10$ to $100\unit{m}$ (the Kolmogorov turbulent cascade dissipation
scale).  One of the most successful approaches to fill this
four-order-of-magnitude wide gap is to assume that the statistics obtained
at observable scales are still valid at smallest scales.  The properties of
the global system, from observable to non-observable scales, can then be
investigated.  This, for example, is the idea underlying Hudson's
(1991)\nocite{hud91} critical power-law slope of $-2$ for the distribution
of flare energies.

\subsection{Statistics from previous observational studies}

\label{sec:statobs}

\paragraph{Distributions of events.}
The measurement of the power-law slope for the lowest energy flares has
indeed been a major goal of coronal physics in the last decade.
\citet{asc00} has summarized the distributions of event energies that were
obtained at wavelengths from X rays to ultraviolet (UV), and for event
energies covering a range of eight orders of magnitude from
$10^{17}\unit{J}$ (``nanoflares'') to $10^{25}\unit{J}$ (``flares''). It
seems --- and it is a statement of \citet{asc00} --- that these
distributions can be matched together to form a unique power-law
distribution of slope $\approx -1.8$.

\paragraph{Spectra.}
Fourier spectra of the radiance fields in UV and X rays have also been
obtained \citep[e.g. ][ from Yohkoh/SXT, NIXT and
EIT]{mart92,gom93a,ben97,ber98}. Most of them display power laws (with
indices of $-1.4$ to $-4$), indicating the probable presence of turbulence.

\paragraph{Structure functions.}
Using transverse structure functions, \citet{abr02} have found that the
photospheric magnetic field in active regions is intermittent (and
intermittency is higher for high activity).  \citet{pat02} have also found
intermittency in the time series of radiance (light curves) of the lines
Ne\,\textsc{viii} $77.04\unit{nm}$ and O\,\textsc{iv} $79.02\unit{nm}$.
Moreover, intermittency has been observed \emph{in-situ} in the solar wind
velocity and magnetic field fluctuations \citep{burl91,mar97}.

\subsection{Aim of this study and selection of the data}

Almost all the statistics of UV observations cited above were done on
radiance images obtained by pass-band filter imaging instruments. Here we
want to obtain statistics of the coronal turbulence and subsequent heating,
and this requires to have access to more direct signatures of the basic
fields of MHD, namely the 3D magnetic and velocity fields. Some statistics
of the photospheric magnetic field have been obtained \citep[cf. ][]{abr02},
but they are limited to the line-of-sight component.  As for coronal
magnetic fields, no measurement has been obtained over a whole solar area.
Concerning the coronal velocity field, to our knowledge no statistical study
has yet been done in the framework of turbulence, although velocity
statistics are very important in the study of turbulence.  To address this
subject, we need to use spectroscopic data, and \sumer \citep{sumer} is
still one of the solar UV spectrographs which has the best performances
currently available.  Furthermore, a spectrograph also allows us to get
``purer'' data in radiance, because the line radiance is obtained by line
fitting which removes the continuum background.

In order to do statistics, we need large amounts of data, and also large
fields with a good spatial resolution if we were to compute Fourier spectra
or structure functions over a wide range of scales (or frequencies). The
resolution of SUMER ($\approx1''$) allows us to reach scales which are
almost the smallest scales which are observable by current UV instruments.
On the other side of the range of scales, scales of the order of the size of
the Sun can be reached by SUMER when rastering the whole image of the Sun,
as has been done in the ``full-Sun'' observation program which is detailed
in Sect.~\ref{sec:observprog}.

We also choose to restrict ourselves to quiet regions of the Sun.  We are
looking for statistical results on the turbulent state of the corona and on
the resulting events of energy dissipation that heat the whole corona,
active \emph{and} quiet, and there is \emph{a priori} no reason to exclude
specific features like the supergranulation pattern or active regions. On
the other hand, the statistical weight of one active region may be too high
compared to the total amount of data we have, and the characteristics of a
particular active region may influence the overall statistics. For this
reason, we exclude data containing active regions from our analysis (but we
do not perform any particular treatment concerning the supergranulation); an
analogous study of active regions should be the subject of a subsequent
work.

\section{The data}
\label{sec:data}

\subsection{Observation program and data set}

\label{sec:observprog}

We use data from a series of 36 observations of the full Sun, done in 1996
by one of us (PL). These data, which are listed in Table~\ref{tab:sumlist},
were obtained by the SUMER spectrograph aboard the SoHO satellite
\citep{sumer}, and taken from the SoHO archive at MEDOC\footnote{The MEDOC
  archive is in public access from \url{http://www.medoc-ias.u-psud.fr}.}.
At this time, most of the Sun was quiet, and active regions were rare. In
each observation of this program, 8 rasters in the east-west direction were
done with slit number 2 ($1''\times 300''$), and they were accumulated in
the north-south direction so as to obtain full-Sun images.  The exposure
time was $3\unit{s}$ and the slit moved (in most cases) by 4 elementary
steps ($1.52''$) in the east-west direction.

Spectra were obtained with detector A of SUMER, but contrary to other
observations with this instrument, these spectra (or parts of spectra) were
not sent to the ground: only 5 parameters of 3 spectral lines were
transmitted for each position on the Sun (\ie position of the slit and pixel
along the slit). These parameters were computed on board by SUMER. During
this process, some information was of course lost, but this allowed us, on
the other hand, to get \emph{spectroscopic} maps of the whole Sun at a high
resolution ($1''\times 1.5''$), with low telemetry use
($8\cdot10^7\unit{bits}$), and quickly (within $9\unit{h}$).

The following parameters (or ``moments'') of spectral line profiles were
obtained\footnote{Spectroscopic data and mean peak of spectral radiances
  (given for quiet Sun) are taken from \citet{cur01}.}:
\begin{itemize}
\item \moment{1} peak spectral radiance, \moment{2} Doppler shift, and
  \moment {3} width of the line \svii (transition
  $2\mathrm{p}^63\mathrm{s}^2\mathrm{S}_{1/2} -
  2\mathrm{p}^63\mathrm{p}^2\mathrm{P}_{3/2}$ at $93.340\unit{nm}$, of mean
  peak of spectral radiance $0.57\unit{W\,m^{-2}sr^{-1}nm^{-1}}$),
\item \moment{4} line radiance (integrated spectral radiance) of the line
  \lyeps (transition $1\mathrm{s}^2\mathrm{S}_{1/2} -
  6\mathrm{p}^2\mathrm{P}_{3/2}$ at $93.780\unit{nm}$, of mean peak of
  spectral radiance $1.07\unit{W\,m^{-2}sr^{-1}nm^{-1}}$),
\item \moment{5} line radiance of the line \svij (transition
  $2\mathrm{p}^63\mathrm{s}^2\mathrm{S}_{1/2} -
  2\mathrm{p}^63\mathrm{p}^2\mathrm{P}_{1/2}$ at $94.455\unit{nm}$, of mean
  peak of spectral radiance $0.29\unit{W\,m^{-2}sr^{-1}nm^{-1}}$)
\end{itemize}
Hereafter the moments will sometimes be referred to by their numbers.

The \lyeps line is emitted at a formation temperature of around
$10\,000\unit{K}$ in the high chromosphere, and the \svi lines are emitted
around $200\,000\unit{K}$ in the transition region between the chromosphere
and the corona.  The width (``moment'' 2) is computed (on board) as the
width of the interval where the line profile is higher than half its peak
values, and the Doppler shift (moment 1) is the position of the center of
this interval. The data are then compressed aboard SoHO (to one byte per
moment per pixel), and decompressed on the ground before being analyzed (see
details of compression in Appendix~\ref{sec:comp}).

\begin{figure}[tp]
  \centering
  \includegraphics[width=\linewidth]{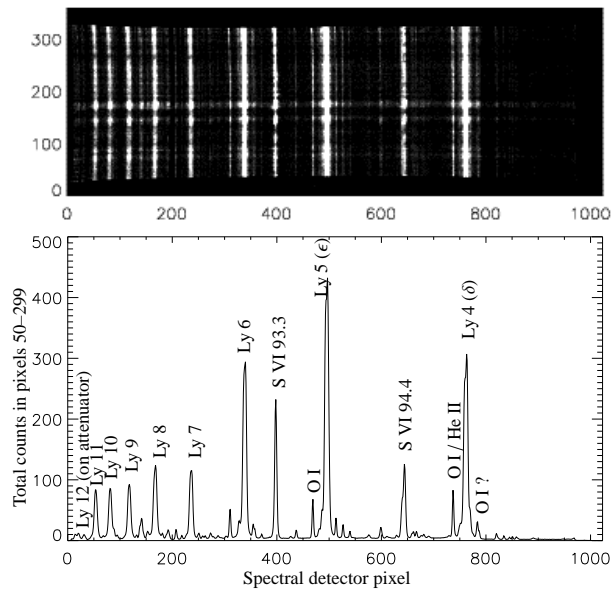}\\
  \vspace{-15mm}
  \resizebox{\linewidth}{!}{%
    \input{spectrum_annotate_en.pstex_t}}
  \caption{Top: context spectrum (full detector A) taken on 4 May 1996 at
    7\,h\,32\,min~UT. Bottom: total counts along
    slit between pixels 50 and 299. The \svi lines and the Lyman series of
    hydrogen (from Ly~4 to Ly~12) can be seen.}
  \label{fig:sumercontext}
\end{figure}

In addition, some full spectra were recorded so as to check for the
positions of the spectral windows (used for the on board computations) and
for the global reliability of the computations. One of these is shown in
Fig.~\ref{fig:sumercontext}. No specific problem was detected. The design of
SUMER, in particular the absence of cosmic rays impacts on the detector,
makes actually the computation of line parameters aboard quite reliable,
notwithstanding the noise problem (see below in Sect.~\ref{sec:noise}).

\begin{table}[tp]
  \centering
  \shorthandoff{:}
  \begin{tabular}{lllll}
    \hline\hline
    Name & Date       & Start    & End      & Notes  \\ \hline
    0401 & 1996/04/01 & 12:30:43 & 17:20:57 &  1\\ 
    0407 & 1996/04/07 & 20:24:54 & 04:56:18 &  \\ 
    0414 & 1996/04/14 & 01:02:49 & 09:34:15 &  \\ 
    0418 & 1996/04/18 & 20:11:39 & 04:43:05 &  \\ 
    0424 & 1996/04/24 & 11:08:45 & 19:40:12 &  \\ 
    0429 & 1996/04/29 & 01:26:38 & 09:58:03 &  \\ 
    0504 & 1996/05/04 & 07:43:09 & 16:14:36 &  \\ 
    0508 & 1996/05/08 & 05:09:03 & 13:40:28 &  \\ 
    0512 & 1996/05/12 & 23:02:03 & 07:33:28 &  AR\\ 
    0517 & 1996/05/17 & 00:47:02 & 08:32:43 &  P1\\ 
    0524 & 1996/05/24 & 08:57:22 & 17:28:48 &  P1\\ 
    0528 & 1996/05/28 & 12:39:04 & 21:10:29 &  P1\\ 
    0603 & 1996/06/03 & 21:13:56 & 05:45:21 &  P1\\ 
    0608 & 1996/06/08 & 21:10:31 & 05:41:56 &  P2\\ 
    0612 & 1996/06/12 & 16:12:47 & 00:44:13 &  P1\\ 
    0616 & 1996/06/16 & 22:41:43 & 07:13:10 &  \\ 
    0621 & 1996/06/21 & 22:48:43 & 07:20:10 &  P3\\ 
    0701 & 1996/07/01 & 19:42:26 & 04:13:48 &  P3\\ 
    0706 & 1996/07/06 & 22:09:36 & 06:41:02 &  AR P2\\ 
    0711 & 1996/07/11 & 17:44:09 & 02:15:35 &  \\ 
    0716 & 1996/07/16 & 19:13:27 & 03:44:49 &  P2\\ 
    0721 & 1996/07/21 & 18:38:19 & 03:09:48 &  \\ 
    0726 & 1996/07/26 & 00:19:04 & 08:50:28 &  \\ 
    0801 & 1996/08/01 & 21:07:56 & 05:39:21 &  AR\\ 
    0806 & 1996/08/06 & 19:14:26 & 00:04:04 &  1\\ 
    0811 & 1996/08/11 & 16:19:29 & 00:50:53 &  AR\\ 
    0814 & 1996/08/14 & 06:07:09 & 14:05:05 &  P3\\ 
    0816 & 1996/08/16 & 20:28:22 & 04:59:45 &  AR P2\\ 
    0821 & 1996/08/21 & 14:41:18 & 23:12:42 &  P2\\ 
    0828 & 1996/08/28 & 09:52:13 & 12:53:48 &  AR P3\\ 
    0902 & 1996/09/02 & 00:11:00 & 08:42:22 &  P3\\ 
    0906 & 1996/09/06 & 04:23:15 & 12:54:41 &  \\ 
    0913 & 1996/09/13 & 21:36:32 & 06:07:58 &  P3\\ 
    0924 & 1996/09/24 & 15:22:43 & 23:53:26 &  AR\\ 
    0930 & 1996/09/30 & 22:02:21 & 06:33:47 &  \\ 
    1005 & 1996/10/05 & 04:42:48 & 13:14:14 &  \\ 
    \hline
  \end{tabular}
  \shorthandon{:}
  \caption{List of the observations of the full Sun done in 1996. The
    4-digit name corresponds to the date of the start of the observation
    (which may end on the following day). Notes: AR, active region near the
    center of the field of view; P1, some pointing problems, or missing
    data, outside of the center of the field of view; P2, these problems may
    affect the center; P3, important problems of pointing or missing data;
    1, the raster step in the east-west direction is $1.14''$ (3
    elementary steps) instead of $1.52''$ (4 elementary steps). }
  \label{tab:sumlist}
\end{table}

\subsection{Correction of the data}
\label{sec:correction}

\begin{figure}[tp]
  \centering
  \resizebox{\linewidth}{!}{%
    \input{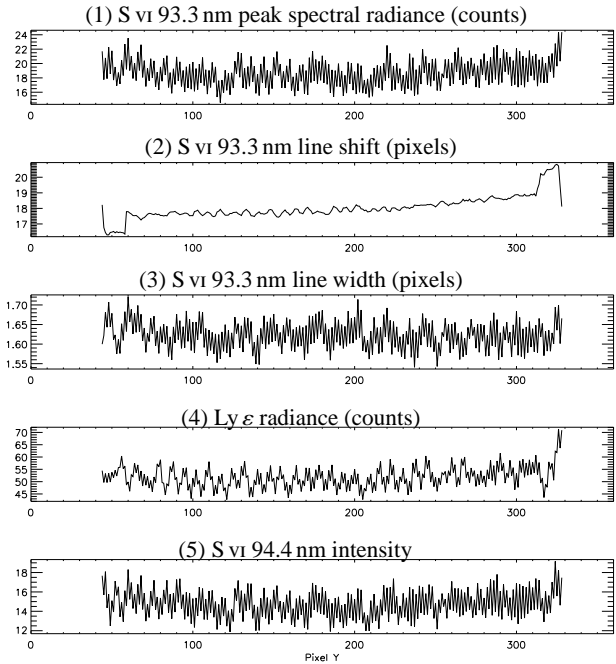}}
  \caption{Average profiles along the slit over all images (excluding
    specific structures), for each of the 5 moments, in counts or pixels. In
    particular, the well-known odd-even pattern of the detector is clearly
    visible. Note that these cuts represent averages around disk center
    only.}
  \label{fig:sumeryprof}
\end{figure}

As all the data do not come with line profiles, the usual procedures of dark
current removal and flatfield correction cannot be applied. Thus we have to
assume that the variations along the slit of averages of data for different
dates and positions result from systematic instrumental effects and not from
solar structures (actually, large structures like active regions were
excluded from the computation of the averages).  These averages, shown in
Fig.~\ref{fig:sumeryprof}, are then subtracted from the raw data (for
Doppler shift) or divide the raw data (for radiances), so as to get
corrected data. For convenience, the units of the resulting data will
hereafter be named ``data units'' and denoted $\unit{du}$.

\subsection{Data values and noise}
\label{sec:noise}

\begin{figure}[tp]
  \centering
  \includegraphics[width=\linewidth]{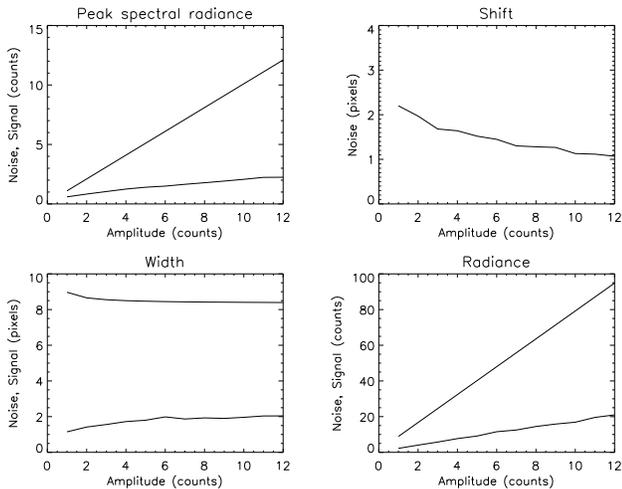}
  \caption{Noise (lower curves) and signal (upper curves, when relevant) in
    each pixel as a function of the amplitude of the Gaussian profile used
    for the Monte Carlo simulations, for peak spectral radiance, line shift,
    line width, and line radiance.}
  \label{fig:sumnoise}
\end{figure}

The context spectra (Fig.~\ref{fig:sumercontext}), which were taken with an
exposure time of $300\unit{s}$, show most line profiles with a few hundreds
of total counts at peak, whereas the full-Sun images were taken with a
$3\unit{s}$ exposure time. The counts on the detector during the full-Sun
observations are therefore low, which is confirmed by the distributions of
the values of radiance moments \moment{1}, \moment{4} and \moment{5} shown
in Fig.~\ref{fig:sumfspres}(d), and noise is an actual issue.  Monte Carlo
simulations were performed to estimate the influence of counting statistics
on the data: we started from Gaussian profiles with background, whose
amplitudes were chosen according to the profiles of the context spectra
($0.1\unit{counts\cdot pixel^{-1}}$ for the background, and up to
$10\unit{counts\cdot pixel^{-1}}$ for the line profiles), and their results
(noise as a function of the line amplitude, for each kind of moment) are
shown in Fig.~\ref{fig:sumnoise}.  The detector noise, of the order of
$10^{-4}\unit{counts\cdot s^{-1}\cdot pixel^{-1}}$, is not taken into
account.  Taking an amplitude of $4\unit{counts\cdot pixel^{-1}}$ as an
example, the noise (standard deviation of the line parameters computed from
the Monte Carlo simulations) is:
\begin{itemize}
\item $1.28\unit{counts\cdot pixel^{-1}}$ for the peak spectral radiance (as
  moment 1),
\item $1.64\unit{pixels}$ for the line shift (as moment 2),
\item $1.77\unit{pixels}$ for the line width (as moment 3),
\item $8.4\unit{counts}$ for the line radiance (as moments 4 and 5).
\end{itemize}
These values, extracted from the plots of Fig.~\ref{fig:sumnoise},
are consistent with the sensitivity predicted during the mission preparation
\citep{wilhelm89}.

\begin{figure}[tp]
  \centering
  \includegraphics[width=\linewidth]{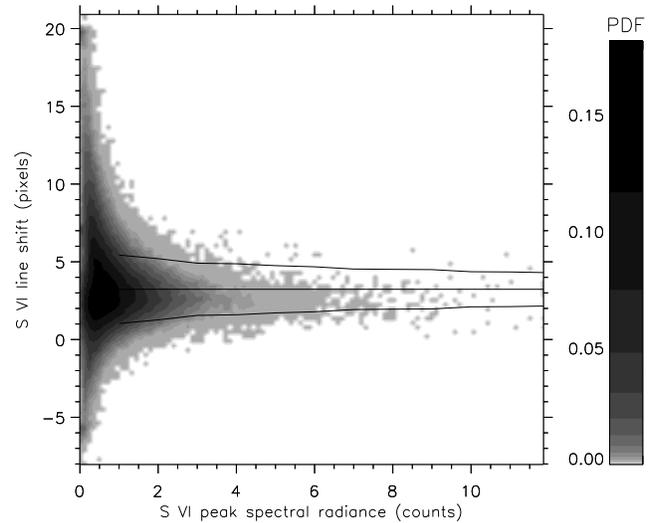}
  \caption{Joint probability distribution function (PDF) of moments 1 (\svii
    peak spectral radiance) and 2 (\svii line shift), in data units (counts
    and pixels). The horizontal line superimposed on this plot is the median
    of moment \moment{2}, and the other curves are the median plus or minus
    the standard deviation of the noise computed by the Monte Carlo
    simulations (Fig.~\ref{fig:sumnoise}), as a function of
    moment~\moment{1}, assuming that the noise in moment \moment{1} is low.}
  \label{fig:ivscatt}
\end{figure}

Among the radiance and line shift data we use, the most acute noise problem
comes from moment~2 (Doppler shift of \svii): as can be seen in
Fig.~\ref{fig:ivscatt}, most of the data are contained in the error bars
coming from a 1-$\sigma$ evaluation of the noise from the Monte-Carlo
simulations, thus noise can account for most of the dispersion of the data
in moment~\moment{2}.  Furthermore this moment is saturated due to the
limited width of the spectral window used to compute the line shift and due
to the compression algorithms (Appendix~\ref{sec:comp}), and this saturation
is still visible in the wings of the distribution of the corrected moment
(as seen in Fig.~\ref{fig:sumfspres}(d.2)); however, this concerns only
$0.5\,\%$ of the pixels and the effect on the statistics is weak.

\subsection{Construction of full-Sun images}

The 8 rasters for each observation are put together, first by using the
pointing coordinates from the headers of the data files, then by a further
adjustment, done so that the limbs on neighboring rasters fit exactly.
Figure~\ref{fig:sumfspres}(a) shows the resulting images for observation
0721.  However, the boundaries between rasters can still be seen, mainly
because the observation times of adjacent pixels in different scans can be
separated by up to 2 hours, and during this time the Sun has rotated by
1.1\,\textdegree\
 in longitude, or $19''$ at disk center as seen
by SoHO. For this reason, when the statistics imply spatial information, we
work with data coming from only one raster at a time. This fact does not
reduce the interest of using these full-Sun images to do statistics, as we
still have data with a large number of pixels produced in a short time.

\newlength{\sunx}
\newlength{\suny}
\makeatletter
\if@twocolumn%
  \setlength{\sunx}{.53\linewidth}%
\else%
  \setlength{\sunx}{.25\linewidth}%
\fi
\makeatother
\setlength{\suny}{1.07\sunx}

\ifendfloat
\begin{sidewaysfigure}
  \begin{tabular}{@{}c*{5}{@{\,}>{\minipage{\sunx}\center}m{\sunx}<{\endcenter\endminipage}}@{}m{0mm}@{}}
    &
    \moment{1} Peak spectral radiance \par \svii &
    \moment{2} Line shift             \par \svii &
    \moment{3} Line width             \par \svii &
    \moment{4} Radiance               \par \lyeps &
    \moment{5} Radiance               \par \svij &
    \tabularnewline[.5em]
    (a) &
    \includegraphics[width=\sunx,height=\suny]{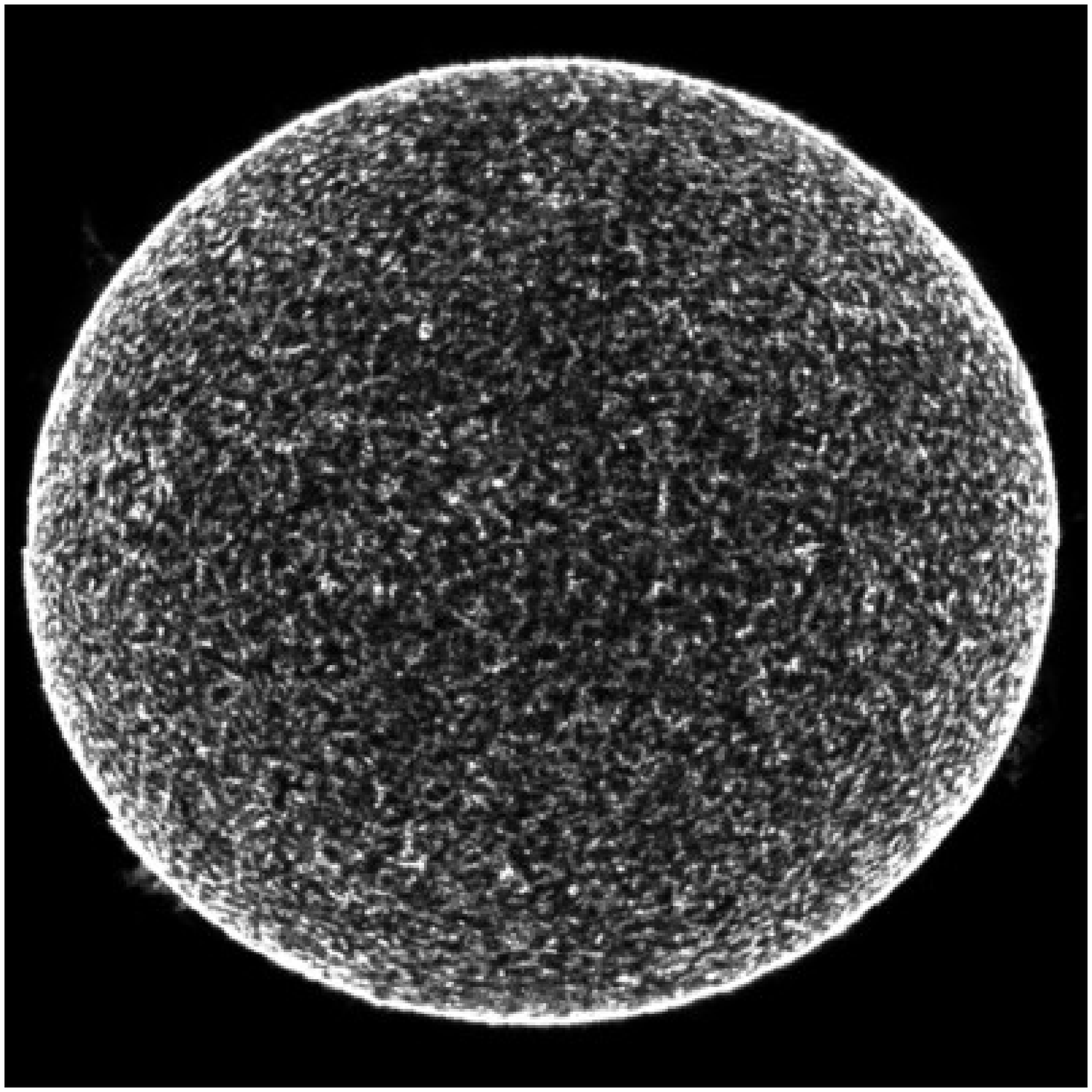} &
    \includegraphics[width=\sunx,height=\suny]{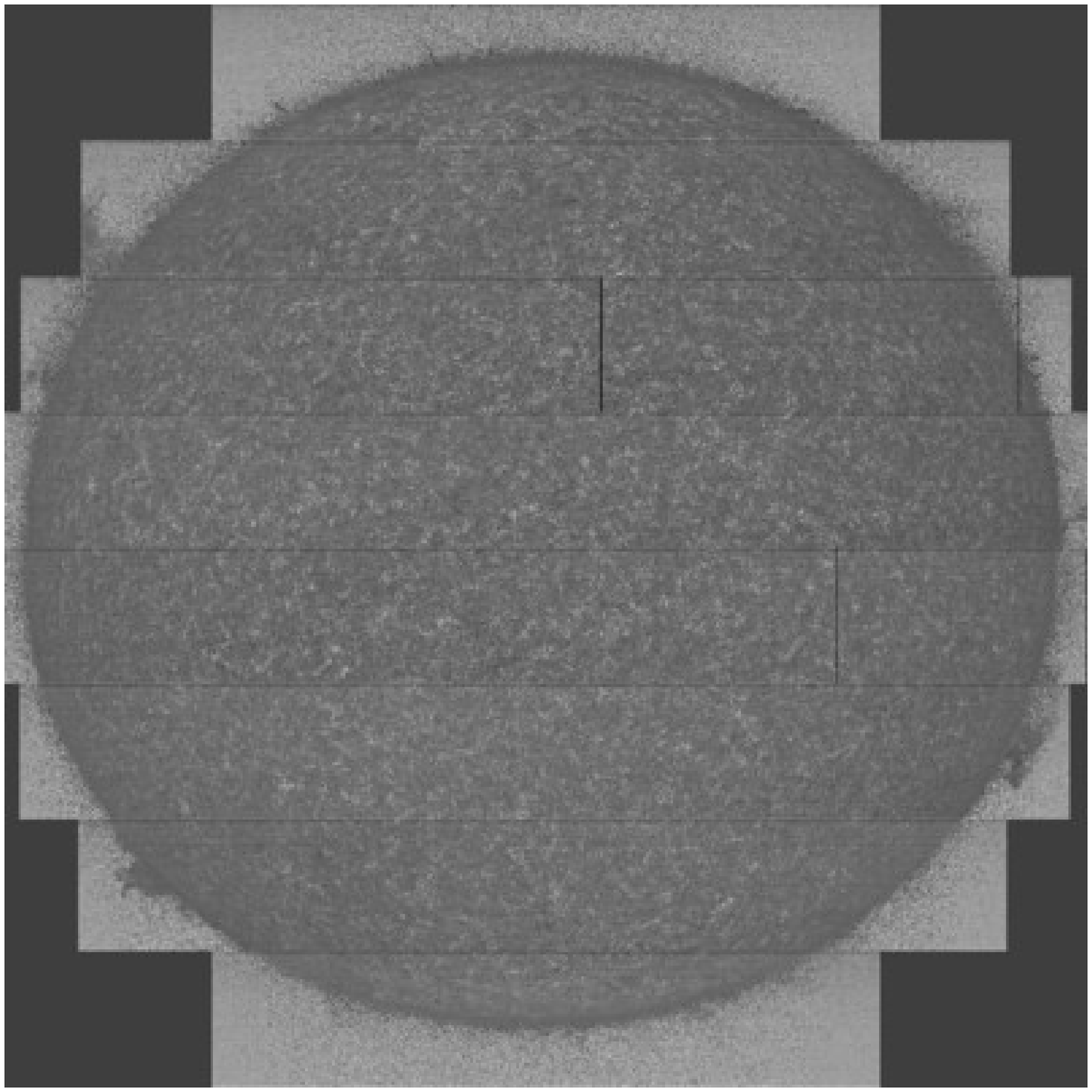} &
    \includegraphics[width=\sunx,height=\suny]{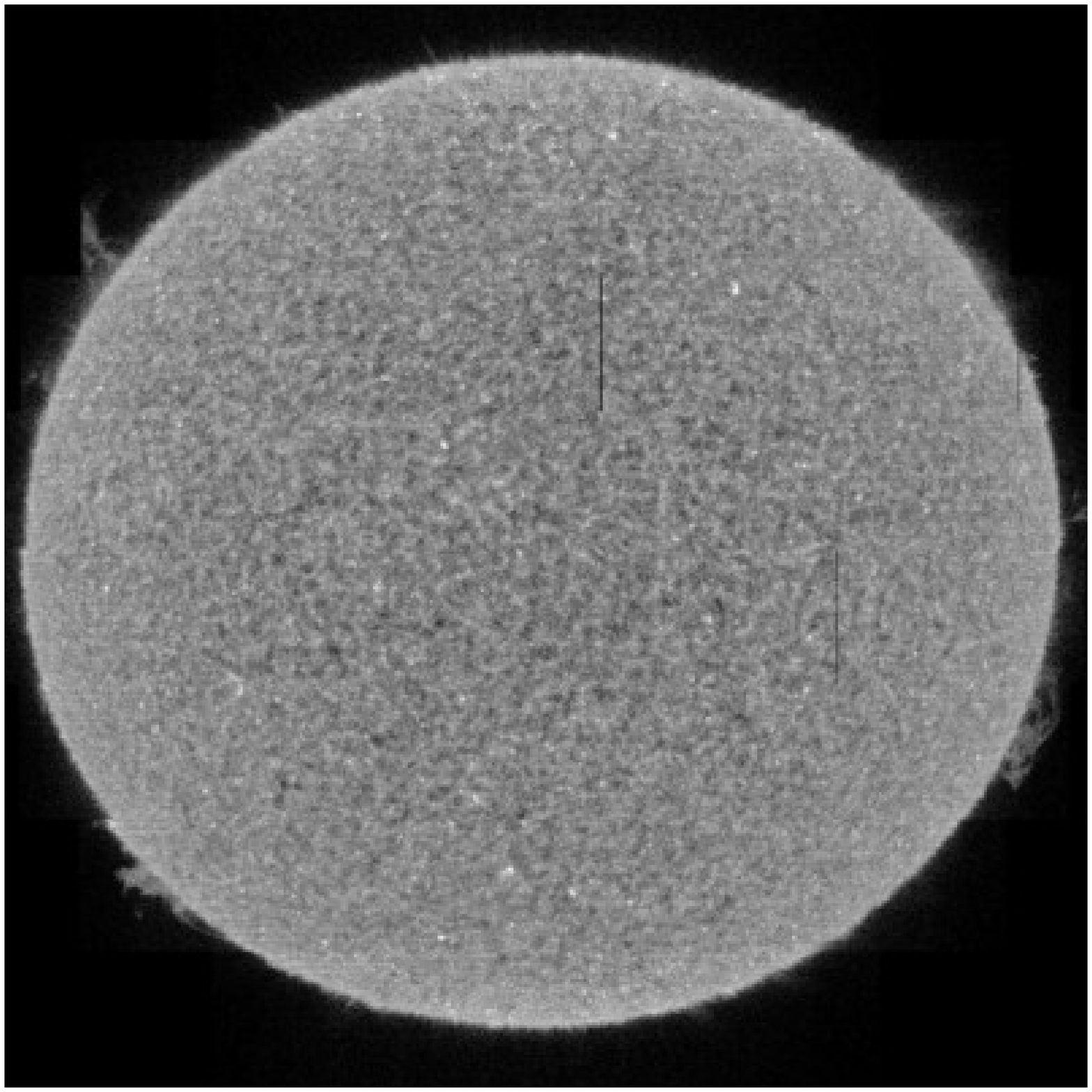} &
    \includegraphics[width=\sunx,height=\suny]{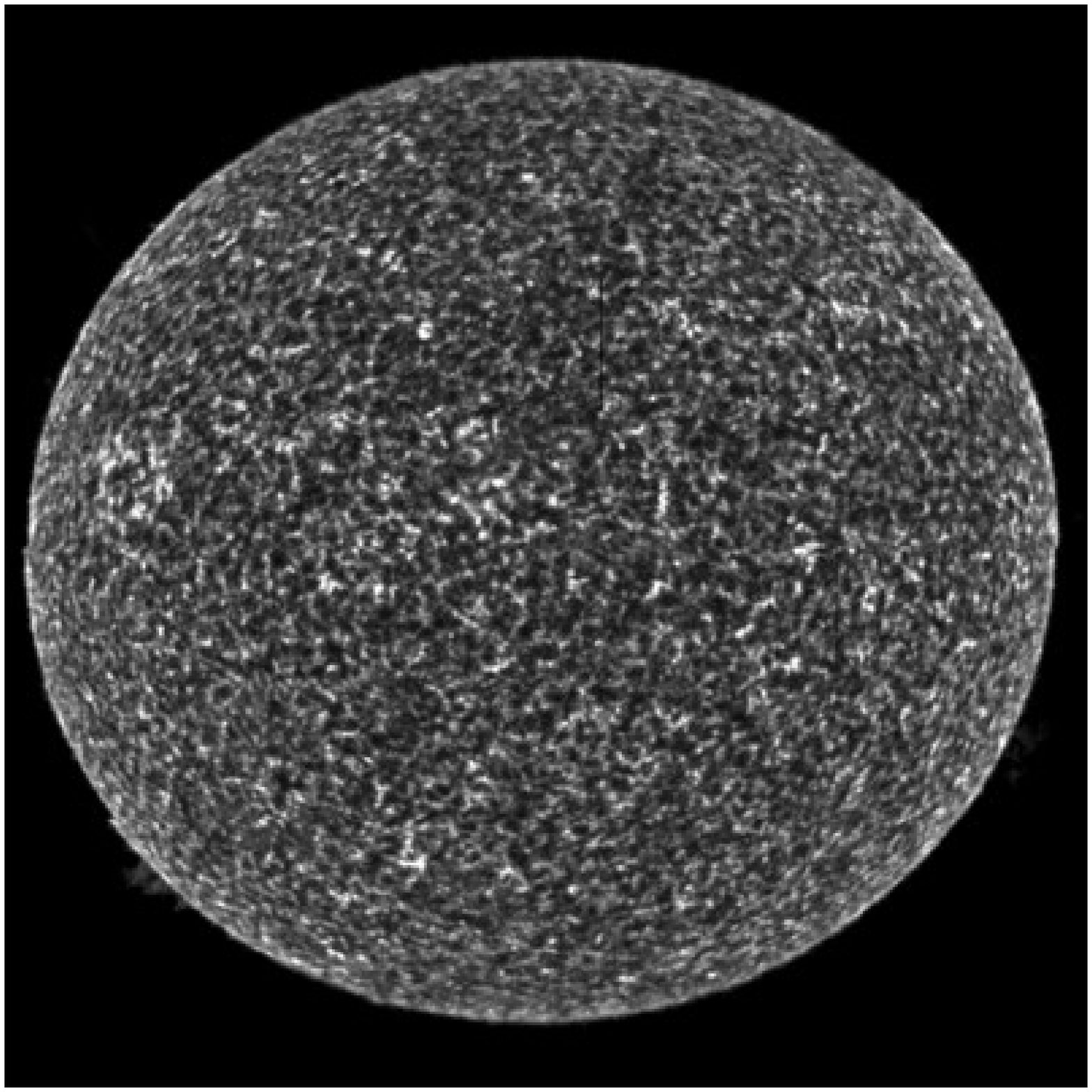} &
    \includegraphics[width=\sunx,height=\suny]{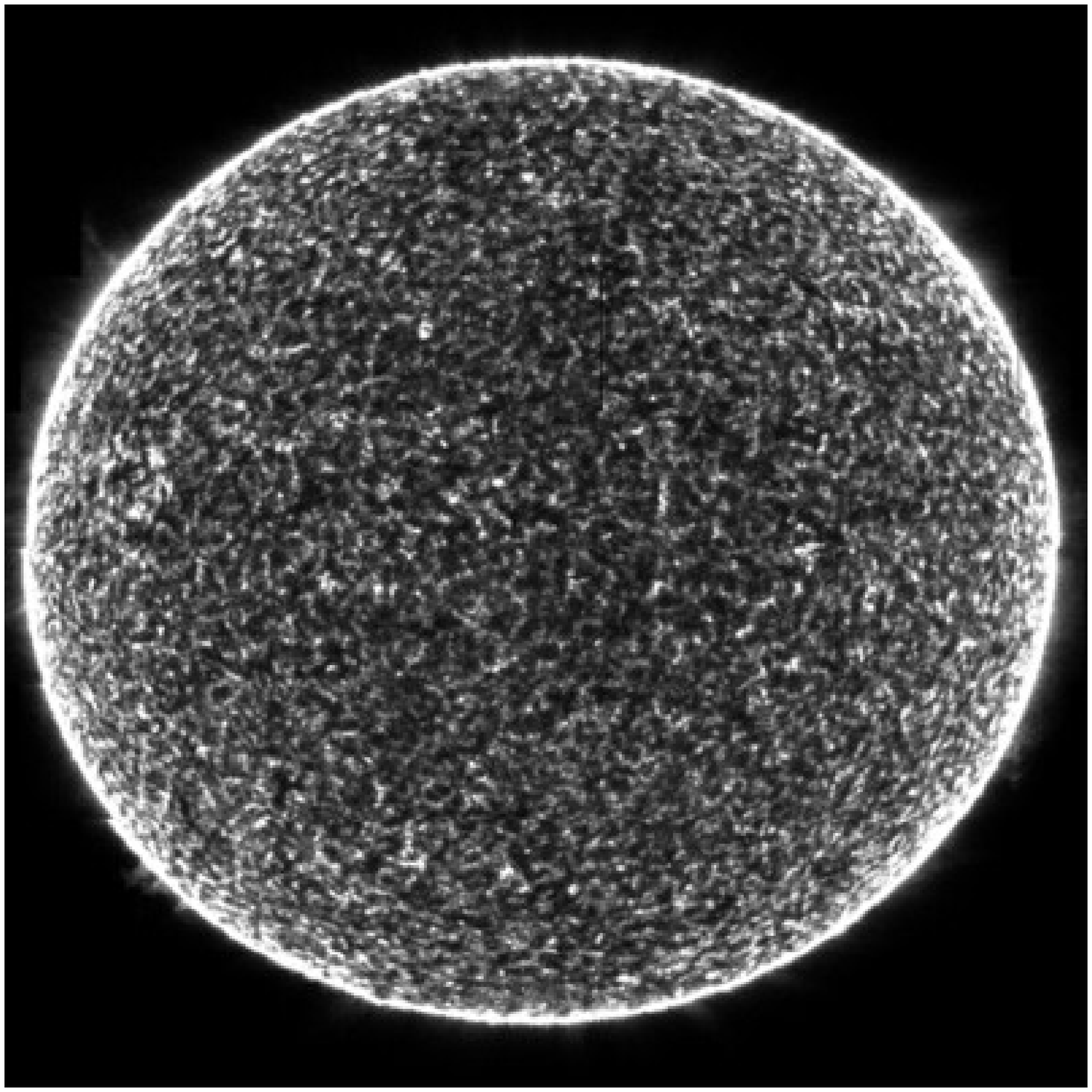} & 
    \tabularnewline[.3em]
    (b) &
    \includegraphics[width=\sunx]{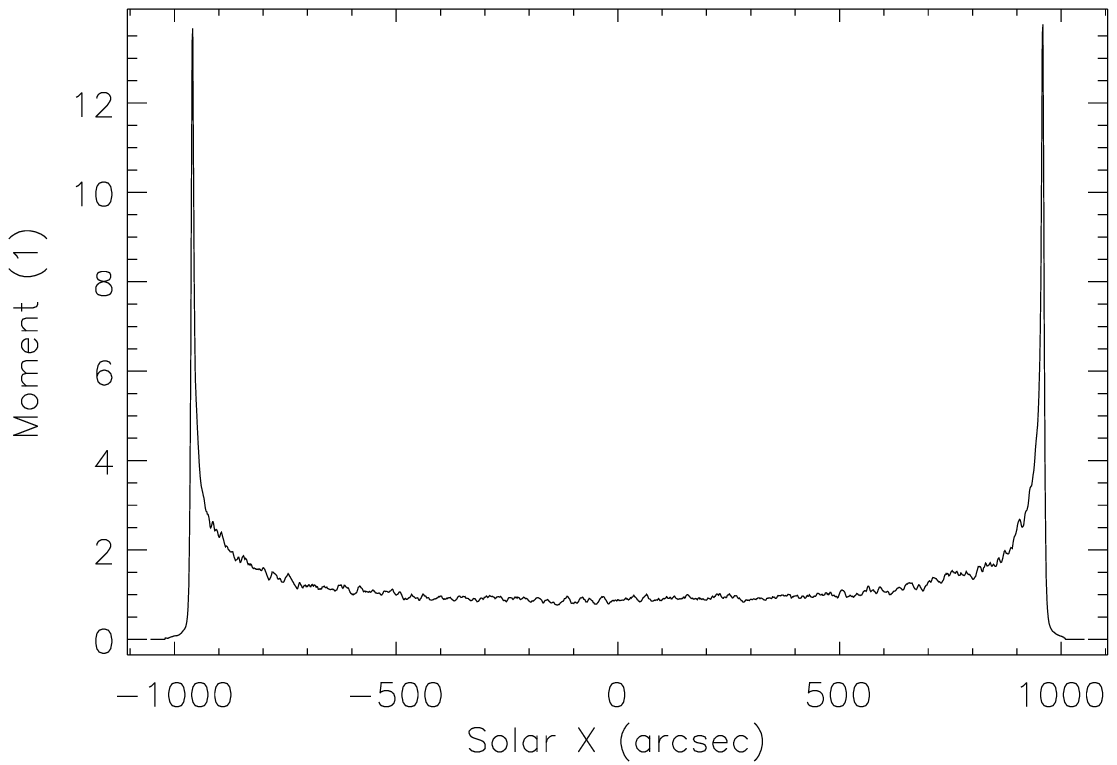} &
    \includegraphics[width=\sunx]{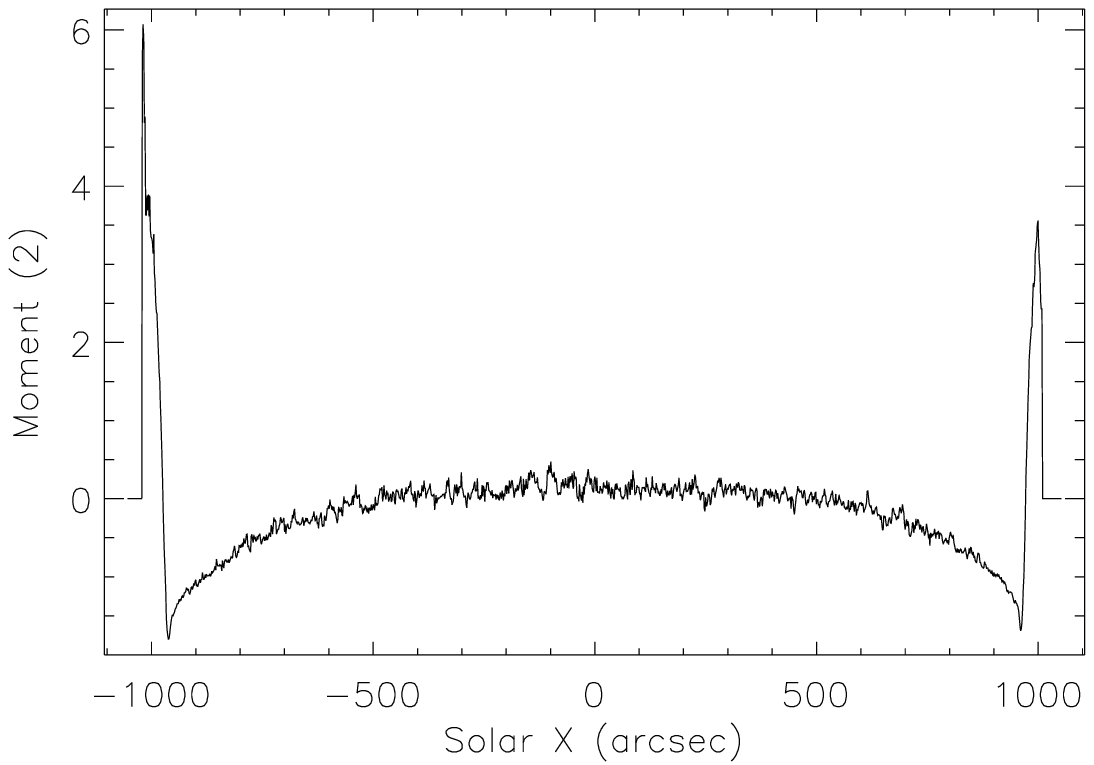} &
    \includegraphics[width=\sunx]{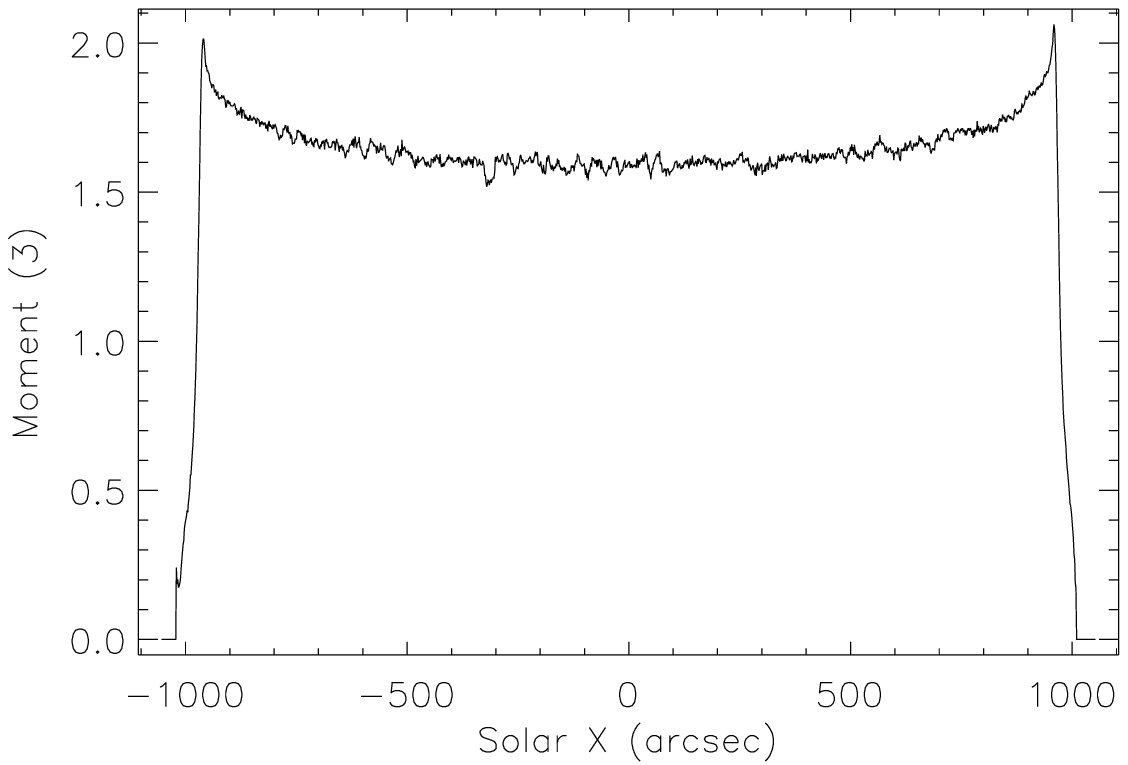} &
    \includegraphics[width=\sunx]{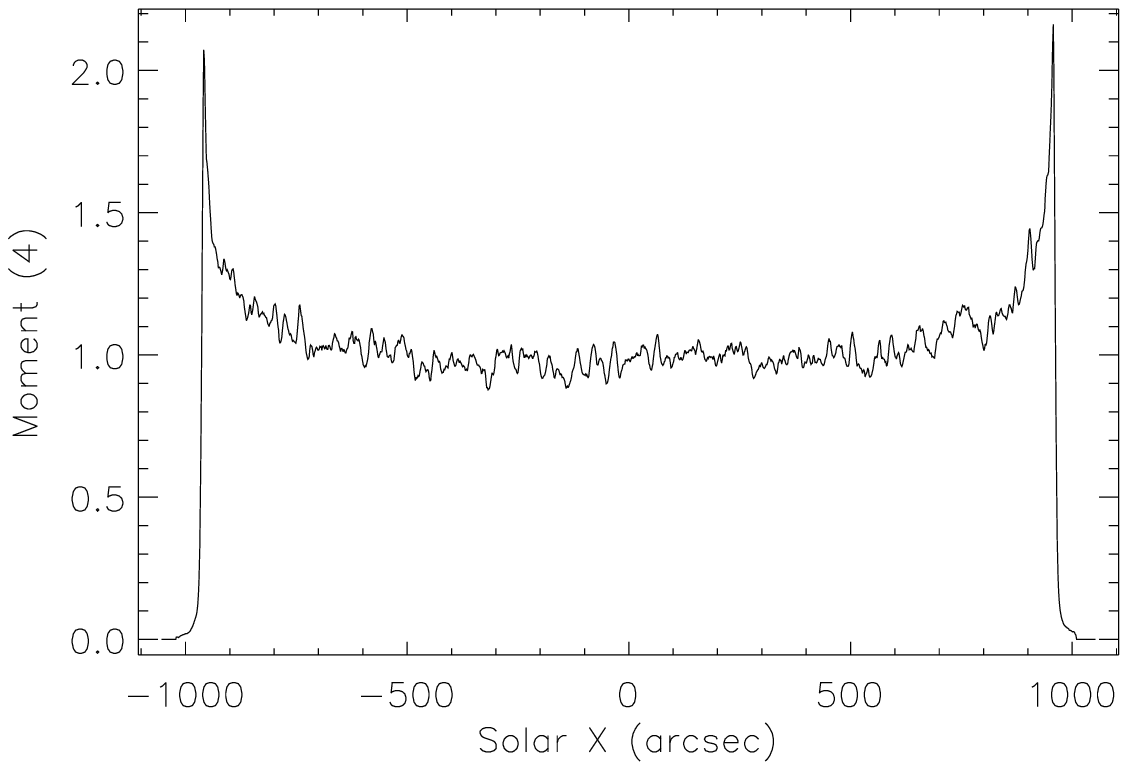} &
    \includegraphics[width=\sunx]{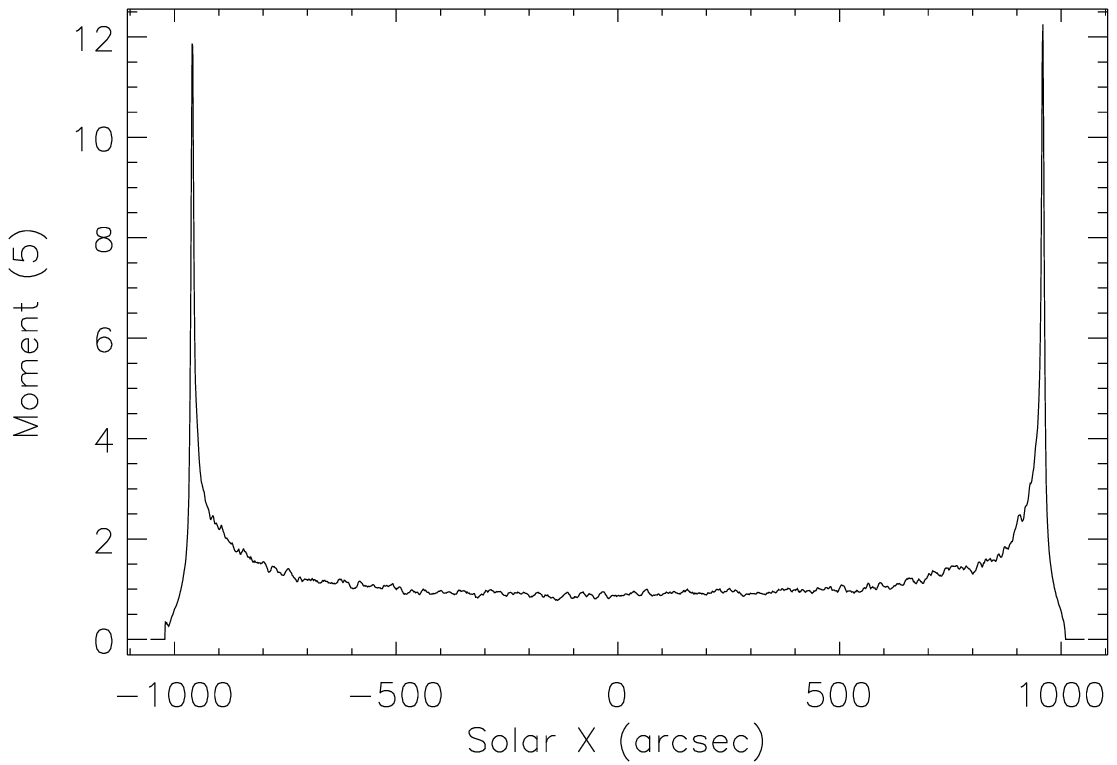} &
    \tabularnewline[.3em]
    (c) &
    \includegraphics[width=\sunx]{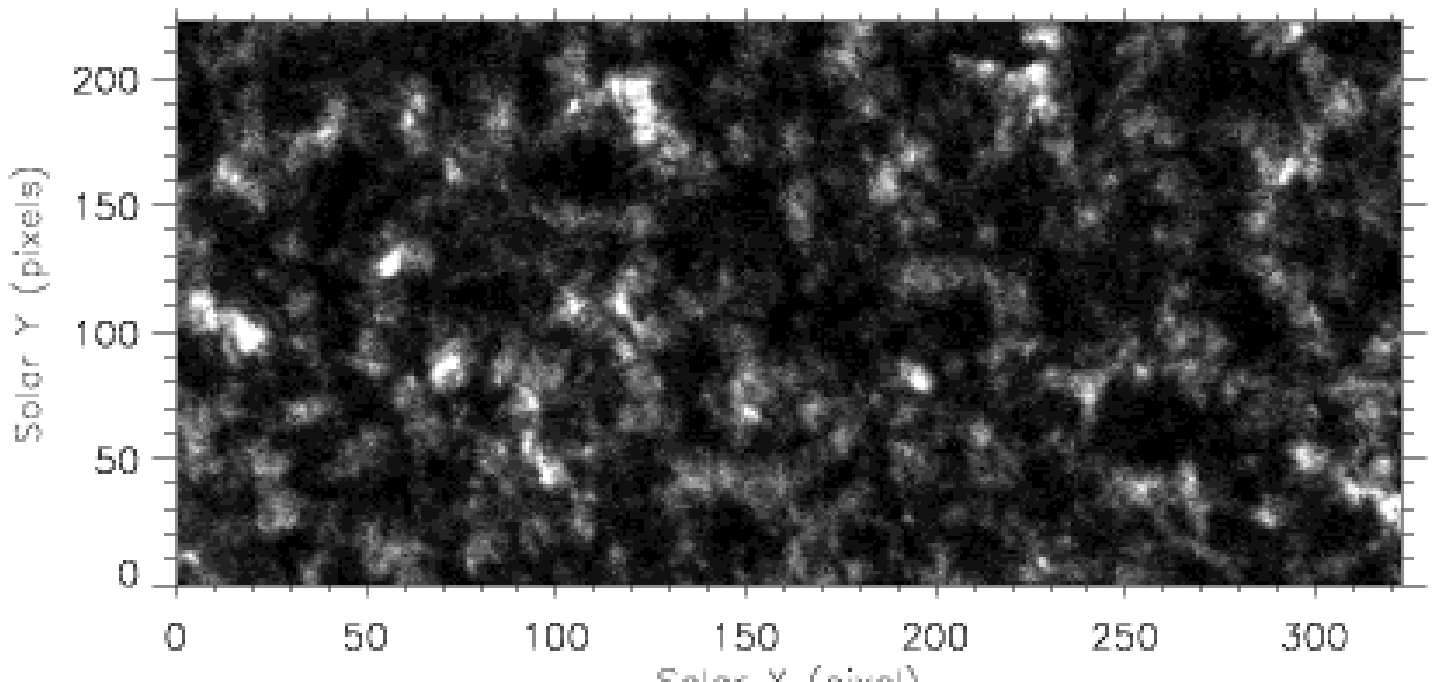} &
    \includegraphics[width=\sunx]{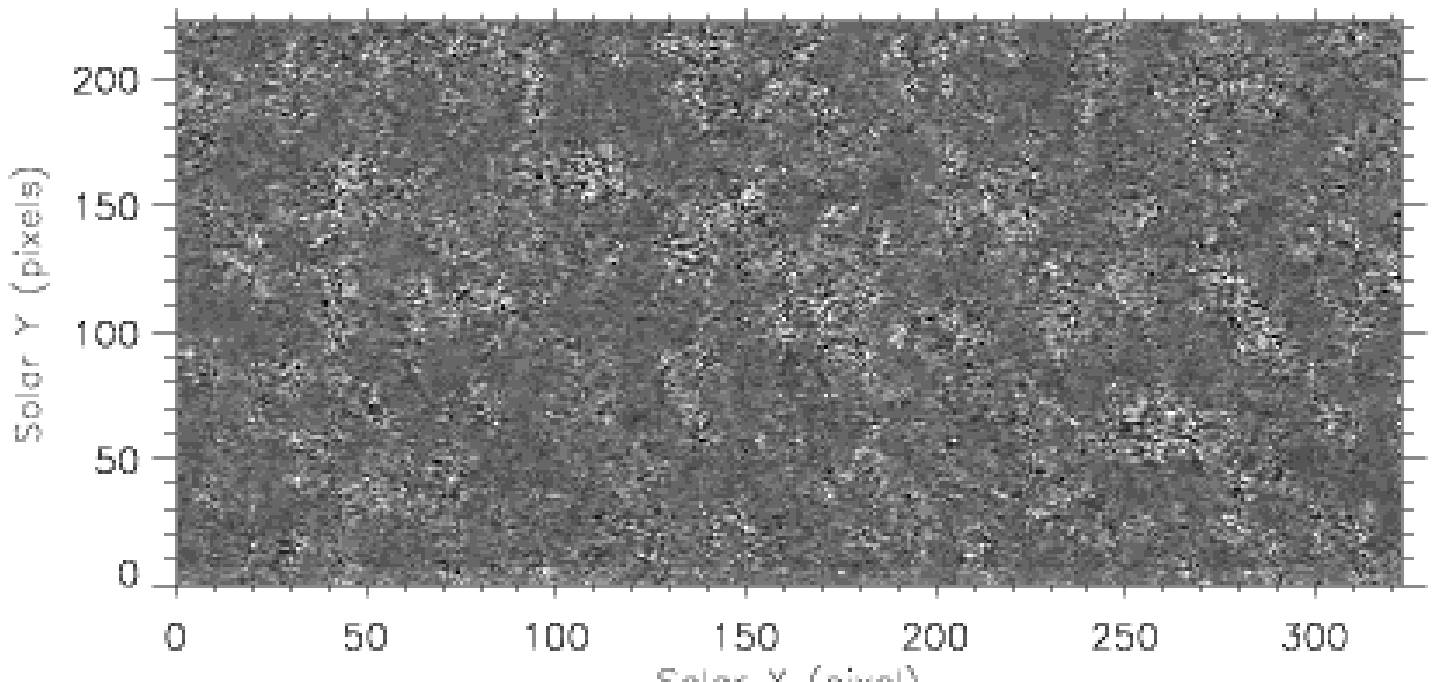} &
    \includegraphics[width=\sunx]{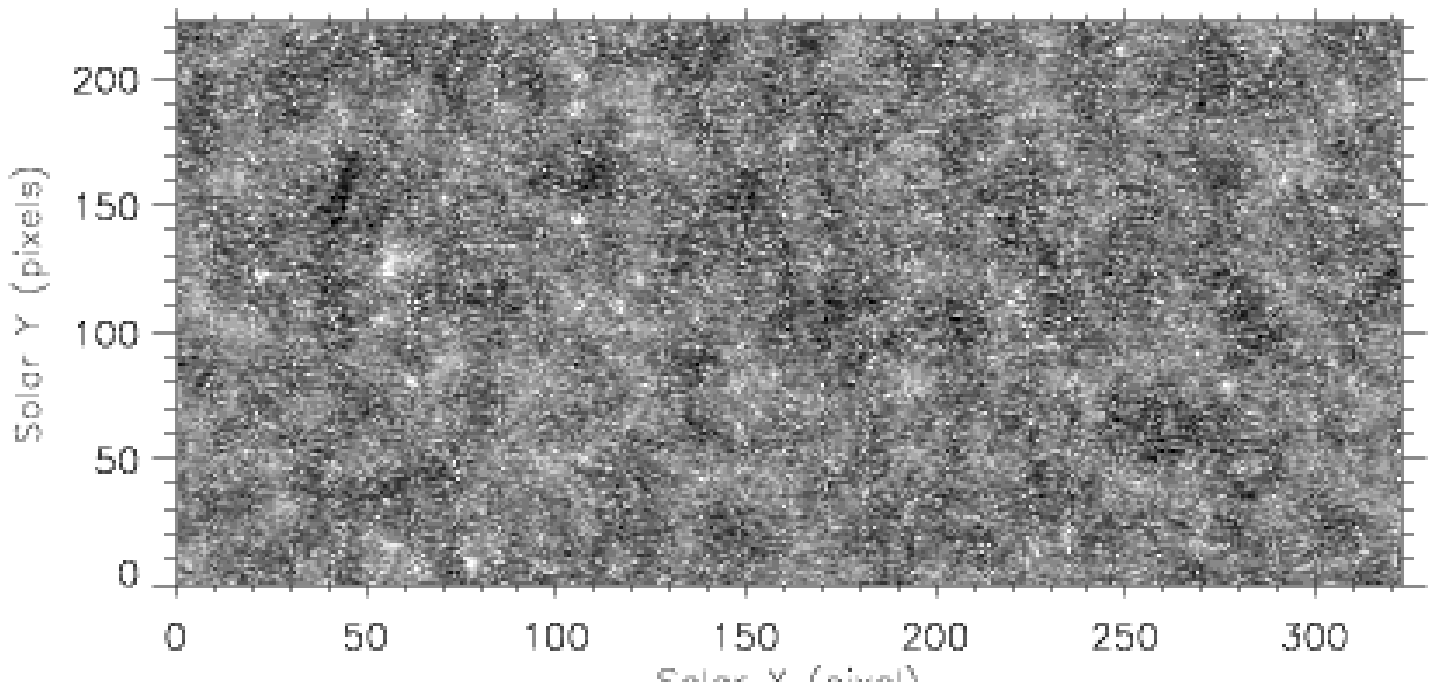} &
    \includegraphics[width=\sunx]{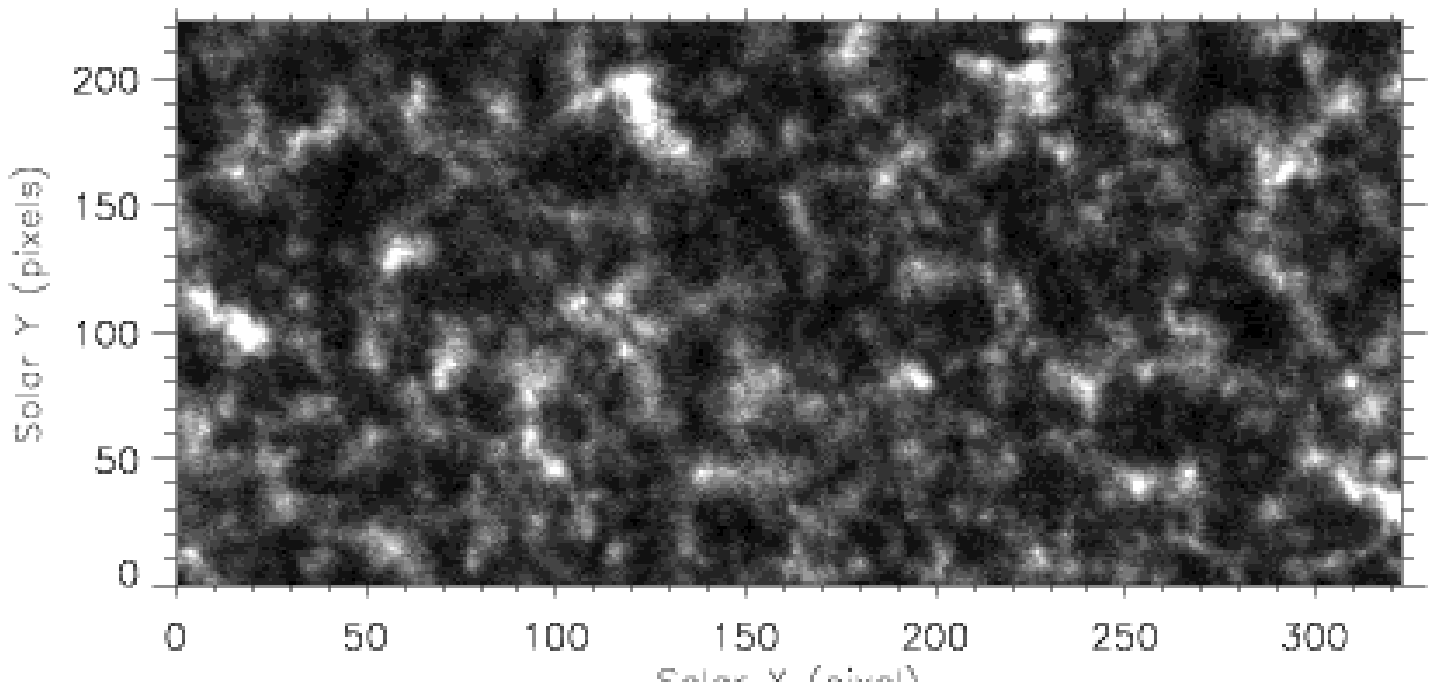} &
    \includegraphics[width=\sunx]{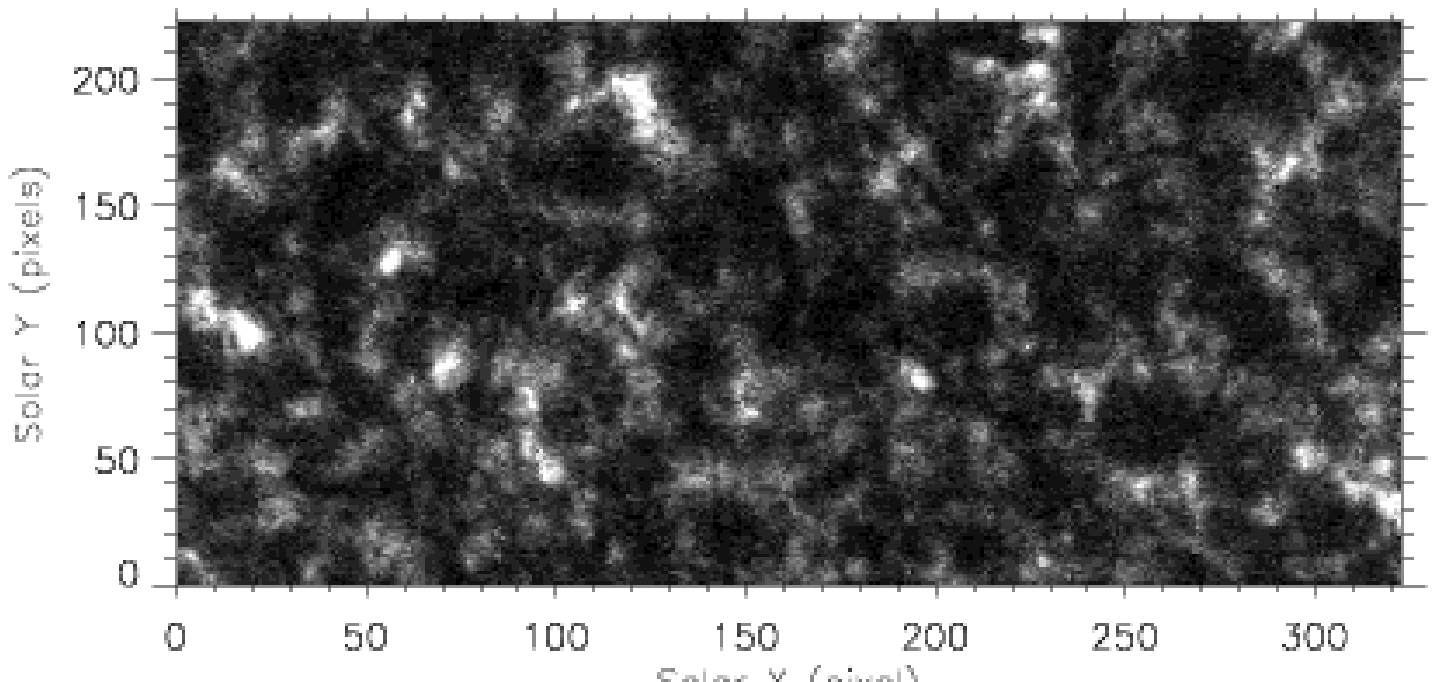} &
    \tabularnewline[.3em]
    (d) &
    \includegraphics[width=\sunx]{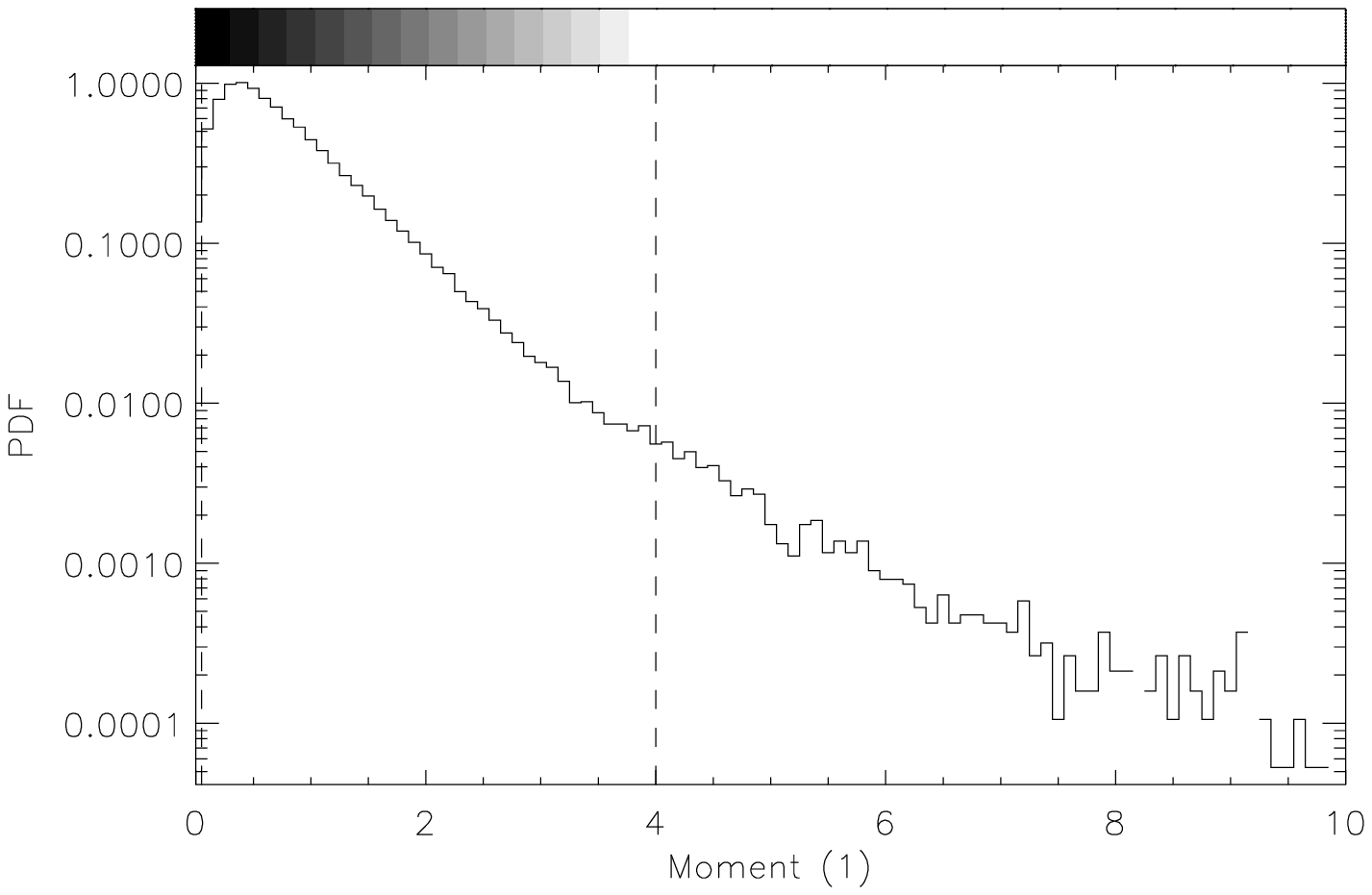} &
    \includegraphics[width=\sunx]{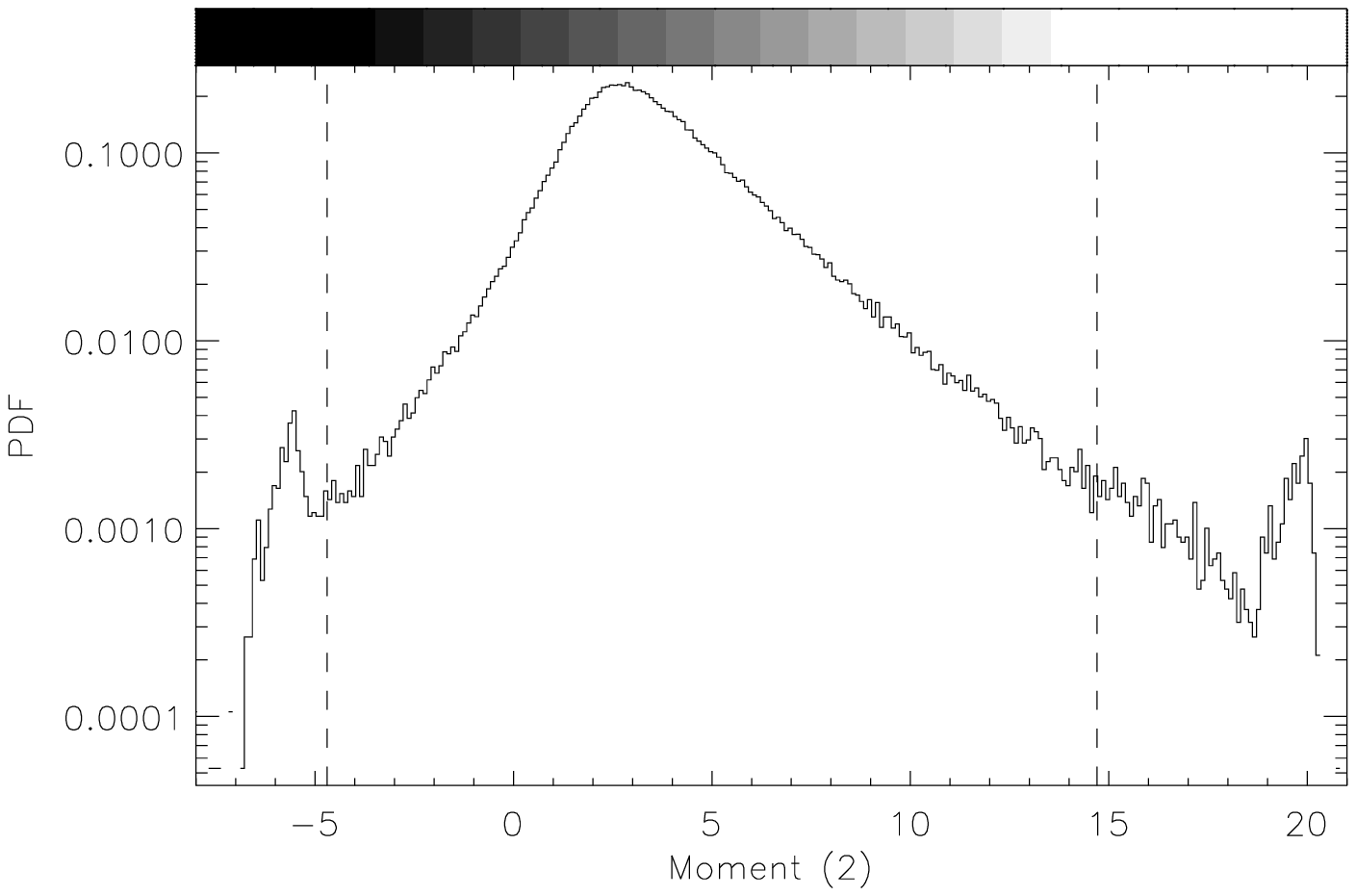} &
    \includegraphics[width=\sunx]{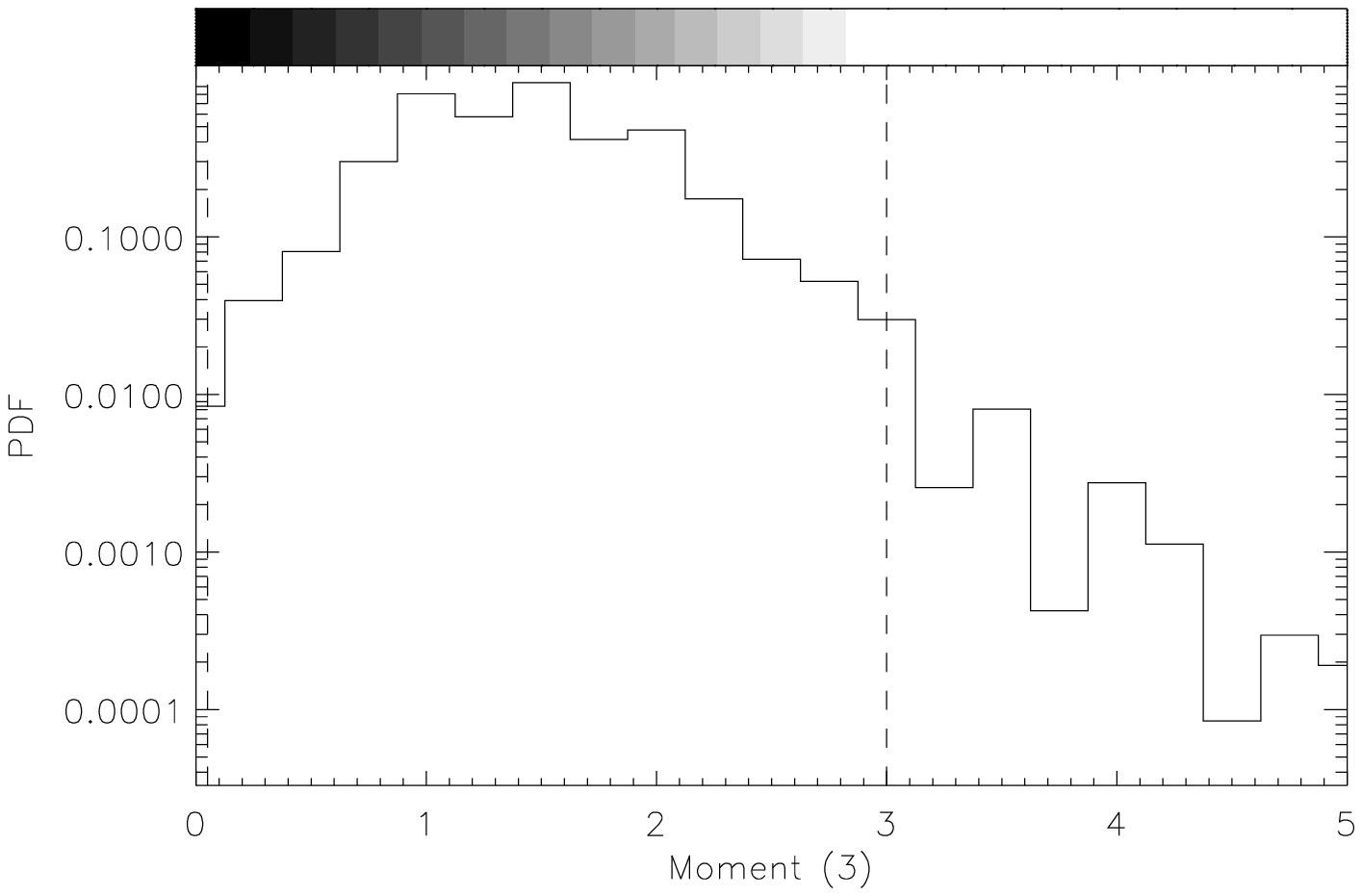} &
    \includegraphics[width=\sunx]{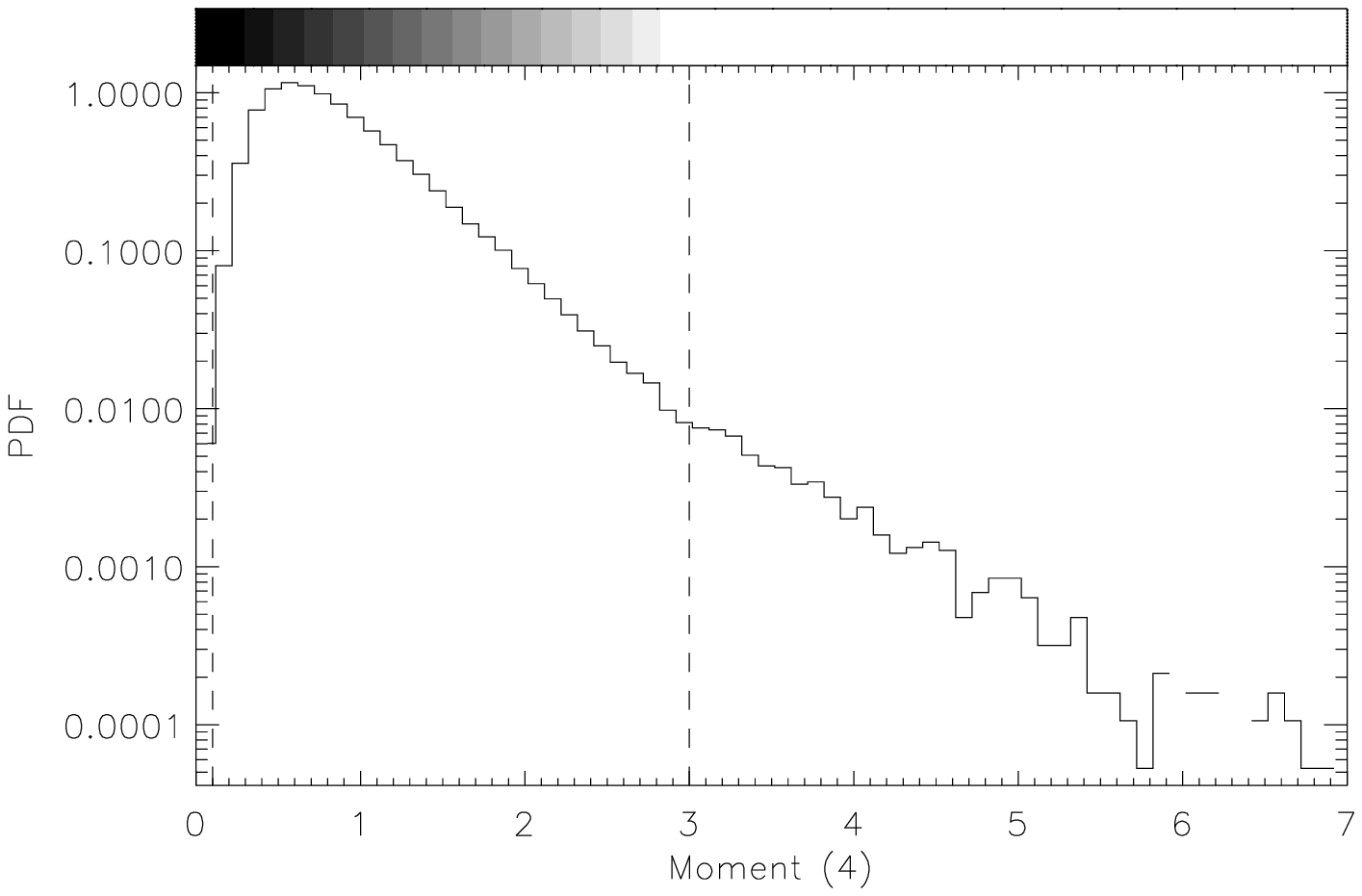} &
    \includegraphics[width=\sunx]{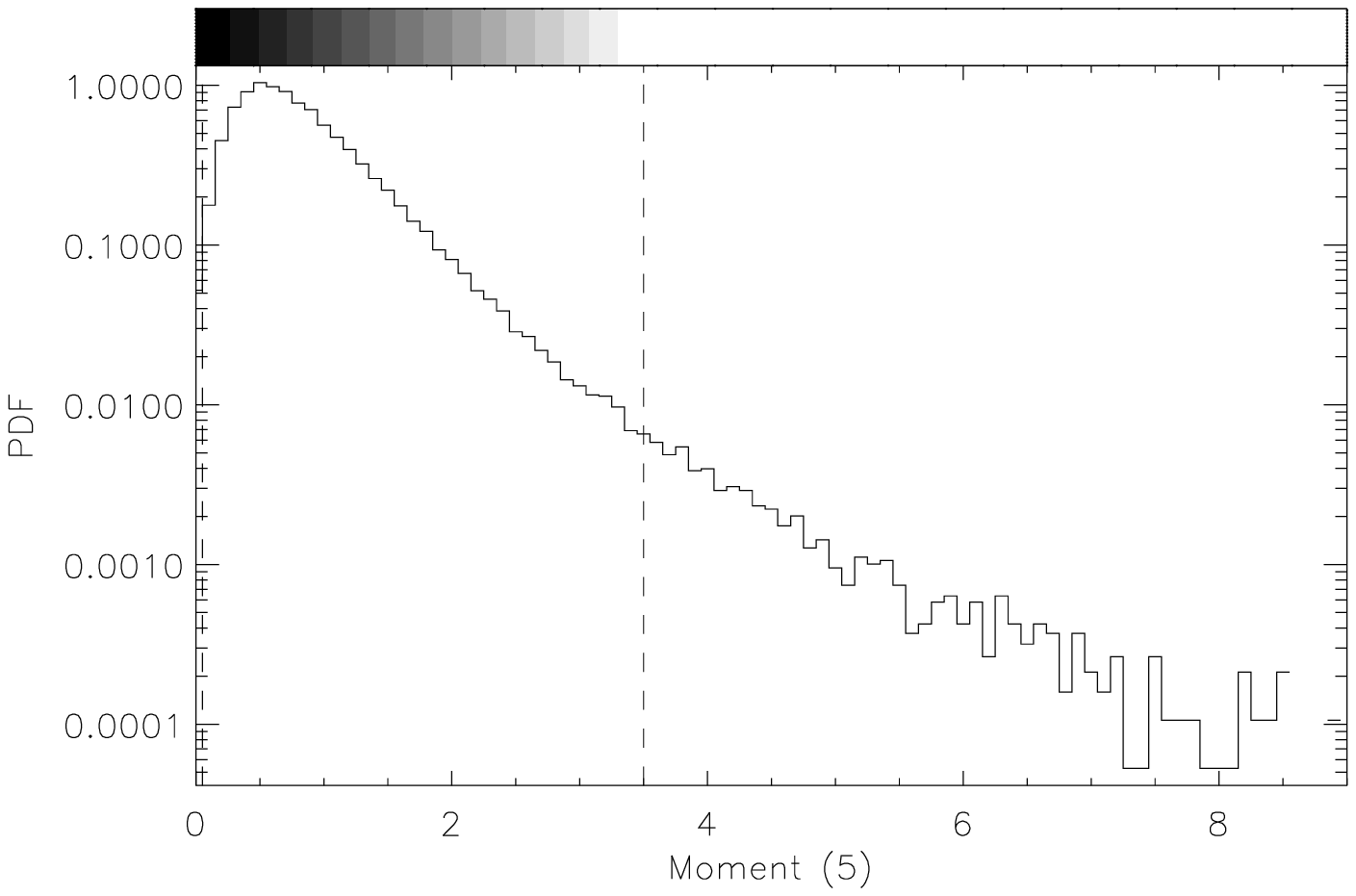} &
  \end{tabular}
  \caption{(a)~Reconstructed images of the full Sun, on 21 July 1996, for
    the 5 moments computed by SUMER aboard SoHO. (b)~Average cuts along the
    equator
    of the full-Sun images, for all five moments. (c)~Detail of the central
    field on 21 July 1996, for all five moments, with an aspect ratio of
    $1$. (d)~Histograms of the values of the moments on a large central
    field on 21 July 1996. The range of the linear grayscale for each moment
    is shown on these histograms and is chosen so that 99\,\% of the pixels
    fall into this range. Note that all the data shown here are the
    moments corrected according to Sect.~\ref{sec:correction}.}
  \label{fig:sumfspres}

\end{sidewaysfigure}
\else
\begin{sidewaysfigure*}
  
\end{sidewaysfigure*}
\fi

\subsection{Profiles of the spectral images along the equator}
\label{sec:calibration}

\begin{figure}[tp]
  \centering
  \includegraphics[width=\linewidth]{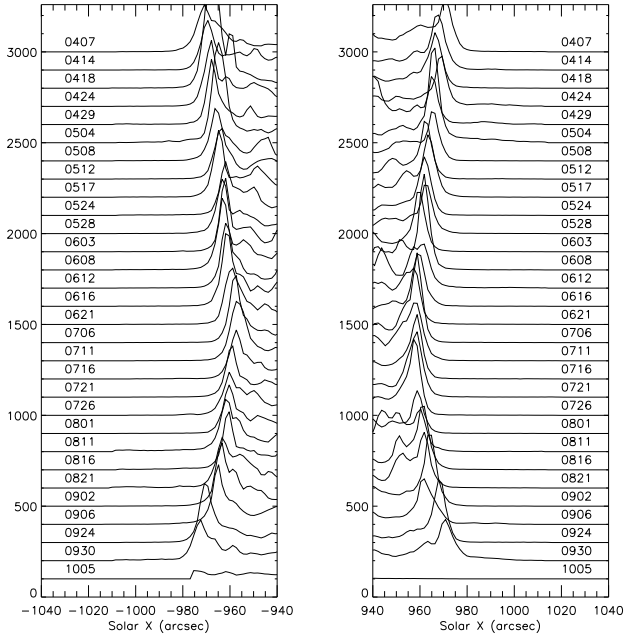}
  \caption{Profiles of \svii radiance in a 100-pixel-wide band around the
    equator, as a function of solar $X$, for all observation dates. The
    units of radiance are data units, and each profile is shifted by 100
    units from the previous one.}
    \label{fig:varsumdist}
\end{figure}

\begin{figure}[tp]
  \centering
  \includegraphics[width=\linewidth]{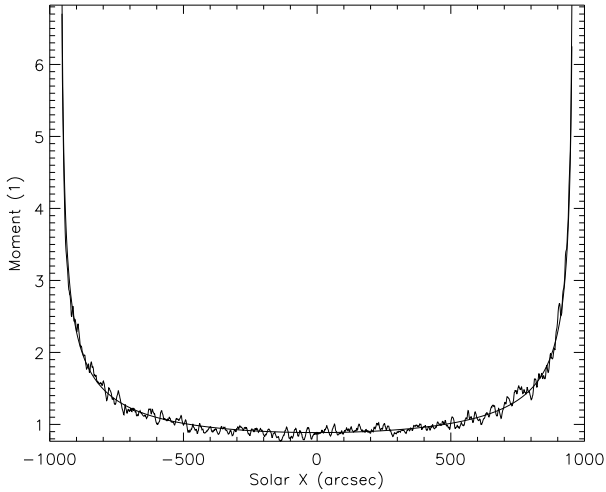}
  \caption{Average profile of moment \moment{1} (\svii peak spectral
    radiance, in data units) along the equator, fitted to the $1/\mu$
    theoretical function.  Pixels at the limb and outside of the limb, where
    the optically thin plan-parallel approximation is not valid anymore,
    have been excluded from the plot}.
  \label{fig:sumiprof}
\end{figure}

We produce averaged profiles of the corrected moments along the equator of
the Sun, in a $200''$-wide band on either side (note that here and hereafter
we approximate the Sun by a sphere and the projection of the solar equator
with the solar disk diameter which is parallel to the $X$ axis).  To do
this, we need to (1) take into account the variation of the apparent size of
the Sun (due to the eccentricity of SoHO's orbit around the Sun), which can
be seen in Fig.~\ref{fig:varsumdist}: the solar $X$ coordinates of the
profiles are normalized to a common solar diameter before computing the
average of the profiles; (2) take into account the curvature of the limb
along the width of the band used to compute the average profiles. Thanks to
this method, the profiles, shown in Fig.~\ref{fig:sumfspres}(b), present
interesting features, like strong limb brightenings in the \svi lines
(moments 1 and 5), which can also be seen in the images of
Fig~\ref{fig:sumfspres}(a). The average profiles of these radiances are
consistent with the theoretical $1/\mu$ limb brightening of optically thin
lines (with $\mu = \cos \alpha$ and $\alpha$ the angle between the normal to
the solar surface and the line of sight). The fit for moment~1 (\svii
spectral radiance) gives $(0.14\pm0.05) + (0.75\pm0.03) / \mu$ with
$\chi^2=5.7\cdot10^{-3}$ and is shown in Fig.~\ref{fig:sumiprof}. A limb
brightening in the \lyeps radiance (moment 4), although weaker because this
line is optically thicker than the \svi lines, has also been measured and
can be seen in Fig.~\ref{fig:sumfspres}(a-b).

\begin{figure}[tp]
  \centering
  \includegraphics[width=\linewidth]{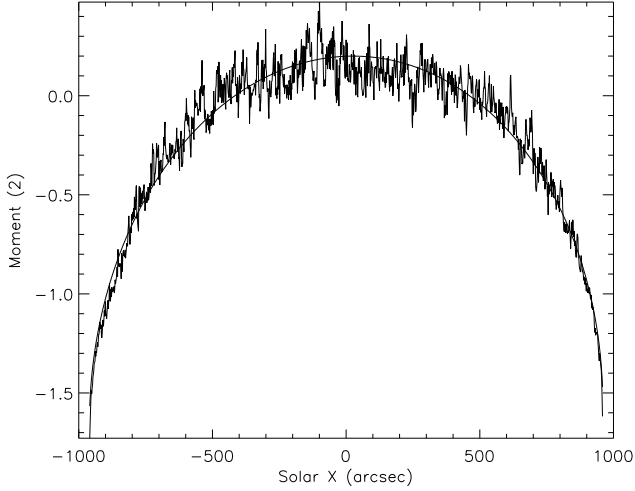}
  \caption{Average profile along the equator of moment~\moment{2} (line
    shift of \svii, in data units), adjusted to the theoretical profile
    given by Eq.~\ref{eq:vlos}.}
  \label{fig:sumvprof}
\end{figure}

The \svii line shift (moment 2), also shown in Fig.~\ref{fig:sumfspres}(b),
presents a curvature that can be interpreted as follows. Let $\theta$ and
$\varphi$ be respectively the heliographic latitude and longitude, with the
origin on the line of sight towards the observer. Assuming that the Sun is a
sphere of radius $R$, rotating at an angular velocity $\Omega(\theta)$
around a north-south axis perpendicular to the line of sight\footnote{We
  consider that the heliographic latitude $B_0$ of the disk center is 0, \ie
  that SoHO is in the equatorial plane; this is an approximation due to the
  angle between this plane and the ecliptic ($7.23^\circ$), and because of
  the excursions of SoHO's orbit outside the ecliptic. But it would only add
  a second-order contribution to the restriction of Eq.~(\ref{eq:vlos}) to
  the equator ($\theta=0$).}, and assuming that matter flows down at a
radial velocity $u(\theta,\varphi)$, the observed velocity, projected on the
line of sight, with positive values for downflows, is:
\begin{eqnarray}
  \label{eq:vlos}
  v_\textrm{los} & = & \Omega(\theta) \cdot R \cos\theta \cdot \sin\varphi 
  + u(\theta,\varphi) \cdot \cos\theta \cdot \cos\varphi \\
  & = & \Omega(\theta) \cdot x + u(\theta,\varphi) \cdot
  \sqrt{1-(x^2+y^2)/R^2} \nonumber
\end{eqnarray}
where $x = R \cos\theta \cdot \sin\varphi$ and $y = R \sin\theta$ are the
heliocentric coordinates. The fit of the profile of moment \moment{2} by
this function (Fig.~\ref{fig:sumvprof}) gives $\Omega=(5 \pm 5) \cdot
10^{-5} \unit{du/''}$ and $u=1.8\pm0.1\unit{du}$ with $\chi^2=7.9\cdot
10^{-3}$, where $\unit{du}$ is the data unit for moment \moment{2}.
Furthermore, a null velocity corresponds to $-1.6\pm0.1\unit{du}$.

On the other hand, the unit of line shift given by the decompression
routines is one pixel on the detector in the direction of spectral
dispersion, \ie $0.0044\unit{nm}$ at a wavelength of $\lambda=94\unit{nm}$,
or a $14\unit{km\,s^{-1}}$ downflow. We would thus expect that
$\Omega=15\cdot 10^{-5}\unit{du/''}$. There is a disagreement with the value
of $\Omega$ obtained previously, indicating that the fit of solar rotation
from average equatorial velocity profiles is perhaps not a good way to
determine the velocity calibration in the case of these observations. The
origin of this problem may be the low signal (see section~\ref{sec:noise}),
which makes it difficult to see the effects of solar rotation (expected to
be $\pm2\unit{km\cdot s^{-1}}$ on the limb). We retain the value
$14\unit{km\,s^{-1}}$ for the velocity unit, which seems to be the most
reliable and which gives physical values for $u$ which are more consistent
with previous observations in the same line
\citep{brekke97,chae98b,peter99b}.

Finally, the \svii line width (moment 3) is also larger near the limb. This
increase is easily explained by an increase at the limb (due to projection)
of the optical depth of this optically thin line: the velocity field of the
matter emitting in this line is averaged over a longer line-of-sight near
the limb than at disk center.

\section{Statistical analysis}
\label{sec:stat}

\subsection{Statistics of  structures}

\begin{figure}[tp]
  \centering
  \includegraphics[width=\linewidth]{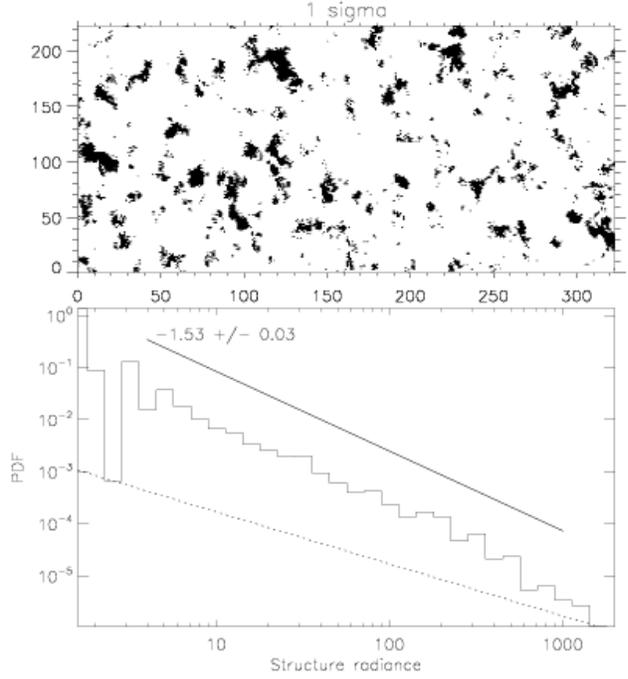}
  \caption{Top: structures (black clusters) determined by a spectral
    radiance threshold of \svii (average plus one standard deviation), on a
    part of the central field, on 21 July 1996.  Bottom: probability
    distribution function (PDF) of the spectral radiance integrated over
    these structures, and its fit to a power law. The dotted line represents
    one event per variable-width histogram bar and is shown as an indication
    of the statistical noise coming from the construction of the histogram.}
  \label{fig:sumthr}
\end{figure}

Bright structures, which are visible in images, can be identified with
energy dissipation events. Here we use the same method as in \citet{ale00}
to define structures: one structure is a set of pixels whose radiance is
above a threshold $L_\mathrm{thr}$, \ie a maximal connex part (cluster) of
the set of image pixels with radiance above the threshold. Such structures,
found using moment~\moment{1} (\svii spectral radiance) on 21 July 1996 and
with a threshold $L_\mathrm{thr}$ set to the average radiance plus one
standard deviation, are outlined on Fig.~\ref{fig:sumthr} (top); 2313 of
them were found in a central area of $700\times270$ pixels.  Since the time
difference between successive images is too long, we cannot use the time
information as in \citet{parn00}.

\begin{figure}[tp]
  \centering
  \includegraphics[width=\linewidth]{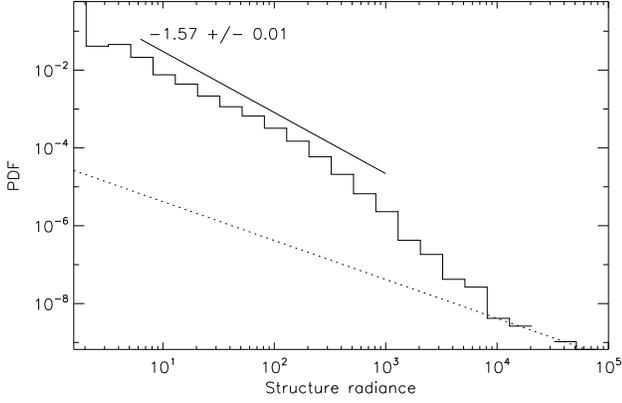}
  \caption{Distribution of the \svii radiance integrated over structures
    defined as in Fig.~\ref{fig:sumthr} and over the 25 best images, and its
    fit to a power law. The dotted line represents one event per
    variable-width histogram bar.}
  \label{fig:sumthrall}
\end{figure}

Let $P_k=\{p_{j,k}, j\in J_k\}$ be the set of pixels in a structure labeled
$k$ ($J_k$ is the set of the indices $j$ of these pixels); the radiance (\eg
moment 1) of the pixel $p_{j,k}$ is called $L_{j,k}$.  The intensity $I_k$
of the event of energy dissipation corresponding to the structure labeled
$k$ is then defined as the integrated radiance on the structure, \ie $I_k =
\sum_{j\in J_k} L_{j,k}$. For the central field of 21 July 1996, we obtain a
power-law distribution\footnote{The probability distribution functions (PDF)
  are obtained by building an histogram and then dividing the height of the
  bars by the total number of elements and by the width of the current bar.}
of index $-1.53\pm0.03$ over more than two decades, as can be seen on the
bottom of Fig.~\ref{fig:sumthr}.  This analysis is repeated on the 25 best
images, and we obtain 40920 structures, whose intensity is distributed as a
power law of index $-1.57\pm0.01$ (Fig.~\ref{fig:sumthrall}). Compared to
the distribution for 21 July 1996, this distribution is extended towards
high intensities because of rare, but intense, events appearing at some
other dates. The results for the other radiance moments are similar and are
summarized in Table~\ref{tab:strucsl}.  The error bars are the uncertainties
on the results of the linear fit of the histograms in logarithmic scale,
taking into acount Poissonian statistics in the counts of events in each
histogram bar. However, they don't take into acount the uncertainties coming
from the selection of the fitting range, which could be evaluated to $0.1$
approximately.

\begin{table}[tp]
  \begin{minipage}{1.0\linewidth}
    \caption{Slopes of power-law distributions of structures found in the
      maps of radiance moments (\svii spectral radiance, \lyeps radiance and
      \svij radiance). Note that the given uncertainties do not take into
      account those coming from the choice of the fitting range (see text).}
    \centering
  \begin{tabular}{rlll}
    \hline\hline
     Obs.       & Moment \moment{1} & Moment \moment{4} & Moment \moment{5} \\
     time(s)    & \svii             & \lyeps            & \svij             \\
     \hline
     On 0721    & $-1.53\pm0.03$    & $-1.46\pm0.03$    & $-1.67\pm0.03$    \\
     On average & $-1.57\pm0.01$    & $-1.44\pm0.01$    & $-1.74\pm0.01$    \\
     \hline
  \end{tabular}
  \label{tab:strucsl}
  \end{minipage}
\end{table}

These slopes are steeper than the slope $-1.2$ found with a similar method
by \citet{ale00} on SoHO/EIT images. On the other hand, they are less steep
than the slopes of the distributions of event energies found by
\citet{asc00} or \citet{parn00} for example, who use a different method:
they use time information (like time clustering to find events), and they
use some hypotheses to derive event energies from their radiance data.  This
derivation is highly non-trivial, as it depends on the relation between
temperature and thermal energy, on the relation between the area and
thickness of the emitting structures, on temperature and density diagnostics
from observations, on the filling factor, etc. Here we have chosen to remain
close to the observable variables, so as to get results independent from the
assumptions needed to derive event energies. We have to keep in mind though
that integrated radiance is not the event energy, and as it is not computed
over the whole spectrum, it is not even the radiative loss during the event.

\subsection{Fourier spectra}

The omnidirectional Fourier spectrum $P(k)$ of a field is obtained by
computing the integral of the 2D Fourier power spectrum
$P_\mathrm{2D}(k_x,k_y)$ of the field (which takes into account the
different resolutions in the $x$ and $y$ directions, and using a cosine
apodization function in both directions) over a ring of radius $k$:
\begin{equation}
  P(k)\ud k = k \ud k \int_0^{2\pi} \ud \theta \; P_\mathrm{2D}(k\cos\theta,
  k\sin\theta)
\end{equation}
In practice, this is done by computing the average of $P_\mathrm{2D}$ over
the grid points in 2D Fourier space that fall in the range $[k - \delta
k / 2, k + \delta k / 2]$ and by multiplying the result by $2\pi k$.

\begin{figure}[tp]
  \centering
  Moment (1)\\[-3mm]\includegraphics[width=.9\linewidth]{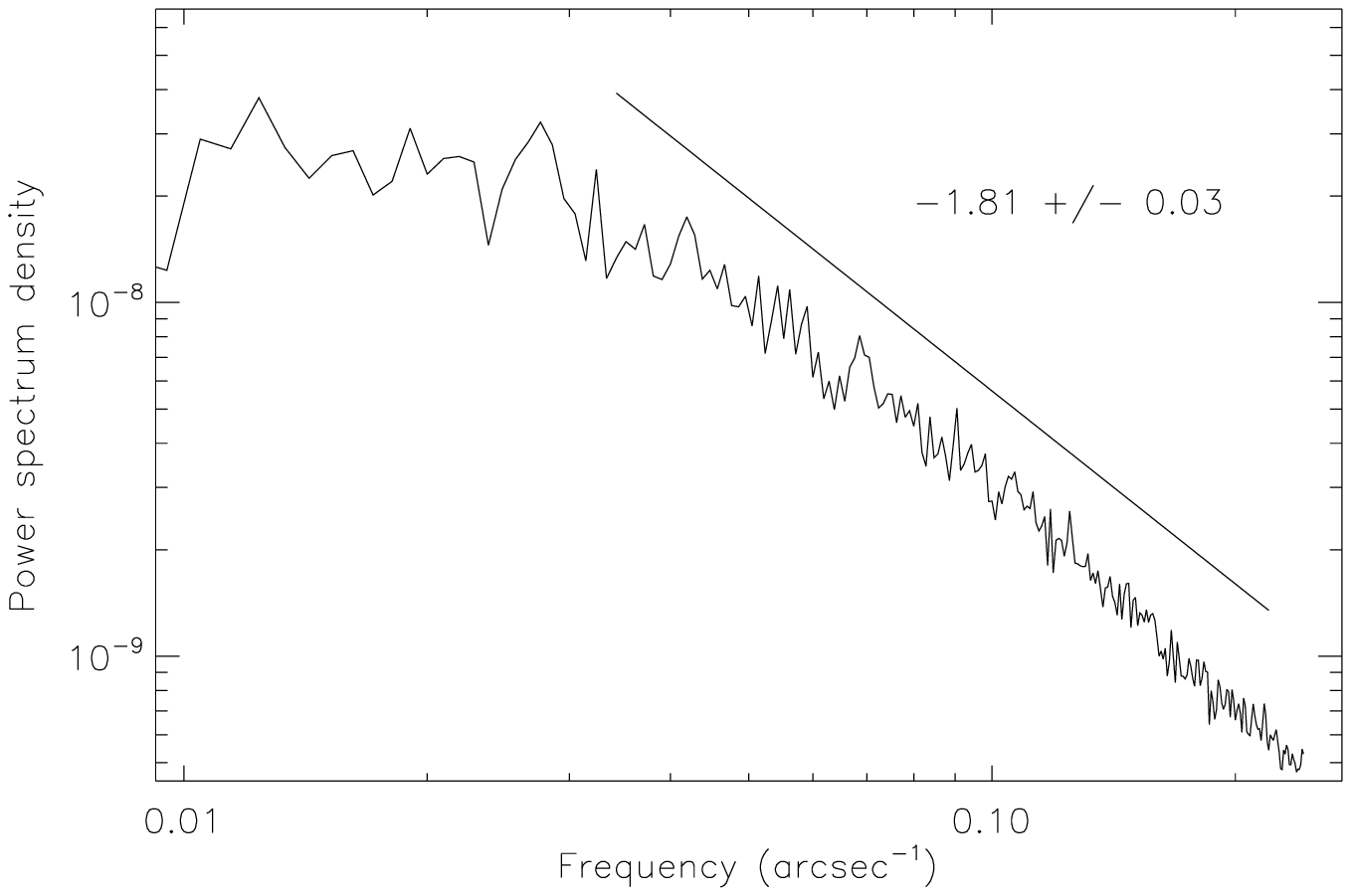}\\
  Moment (2)\\[-3mm]\includegraphics[width=.9\linewidth]{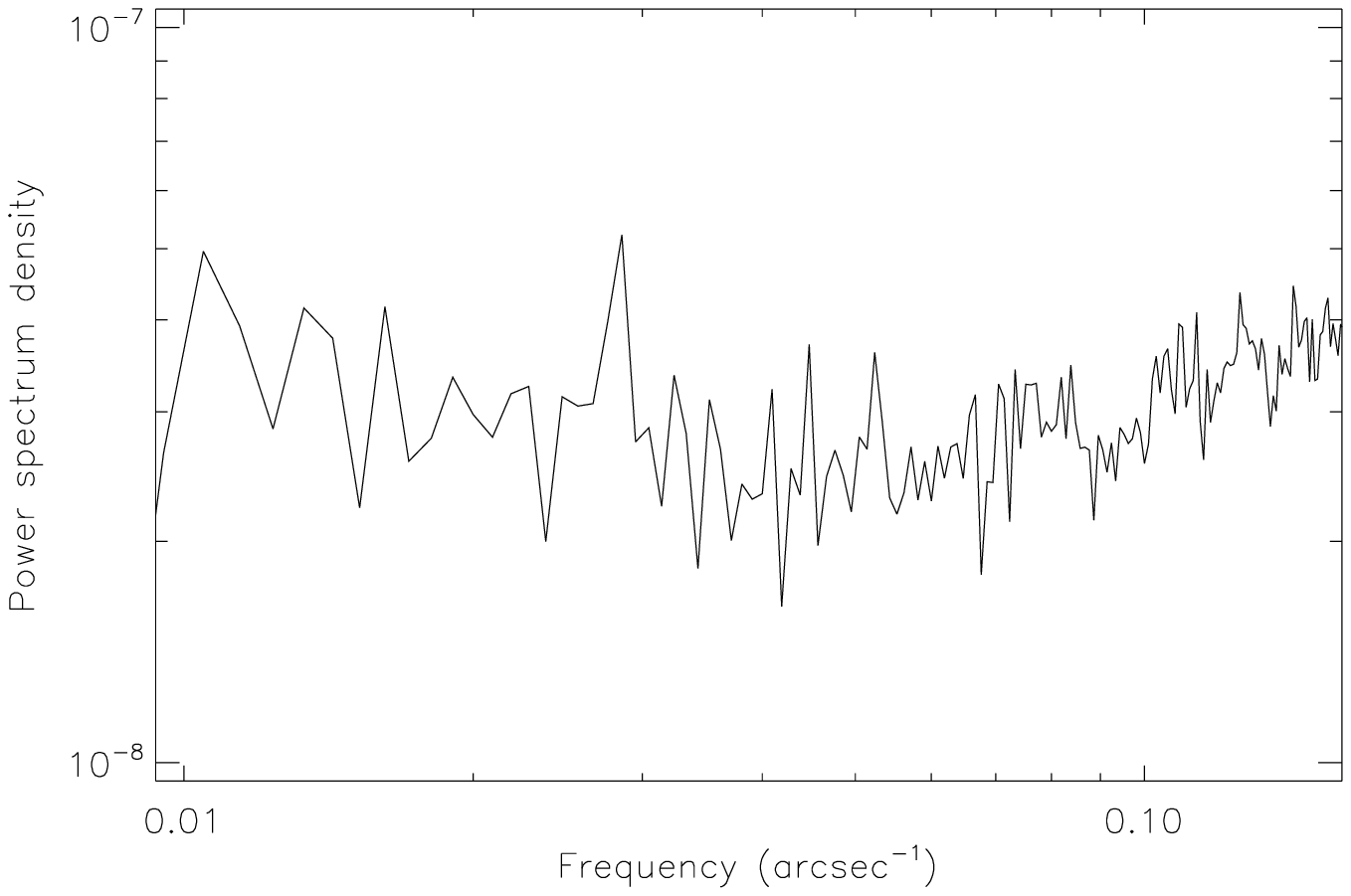}
  \caption{Spatial Fourier spectra of moments 1 (\svii spectral radiance) and 2
    (\svii line shift), on a central field of the Sun, on 21 July
    1996.}
  \label{fig:sumivspec}
\end{figure}

\paragraph{\svii spectral radiance.}
Figure~\ref{fig:sumivspec}.1 shows the spatial Fourier spectrum of the field
of moment 1 (\svii spectral radiance) as a function of the spatial frequency
$f=k/2\pi$. This spectrum is fitted to a power law of index $-1.81\pm0.03$
for scales lower than $25\unit{Mm}$ (corresponding to a $0.03''^{-1}$
spatial frequency), where a break occurs. This scale is at the order of the
scale of the supergranulation, and it is the same scale as the break in the
Yohkoh/SXT X-ray radiance Fourier spectrum obtained by \citet{ben97},
although the slopes of both parts of the spectrum are different there than
in the case of our observations.

\paragraph{\svii line shift.}
The power-law tail of the spectrum of the radiance field indicates that
small scales exist, and they may be produced by turbulence. However, the
direct measurement of the spectrum of the line-of-sight velocity (or line
shift of \svii) --- a physical parameter easier to compare to spectra
predicted by theories of turbulence --- does not yield satisfying results:
these spectra are flat, as can be seen from Fig.~\ref{fig:sumivspec}.2 for
21 July 1996, implying that the noise in the maps of this parameter (moment
2) cannot be ignored.

\subsection{Structure functions}

For a scalar field $a(\vec x)$, we define the increments $\delta_{\vec \ell}
a (\vec x) \equiv a(\vec x +\vec \ell) - a(\vec x)$.  The structure function
$S^q(\ell)$ of index $q$ is the $q$-th moment of the increments $\delta_\ell
a$ of the field at scale $\ell$, \ie:
\begin{equation}
  \label{eq:sfdef}
  S^q(\ell) \equiv
  \left\langle\left|
      \delta_{\vec \ell} a(\vec x)
    \right|^q
  \right\rangle_{\vec{x}, \,\|\vec{\ell}\|=\ell}
\end{equation}
where $\vec x$ and $\vec x + \vec \ell$ go through all the field at a given
$\|\vec \ell\|$. The normalized structure functions are:
\begin{equation}
  \label{eq:sfndef}
  S_\textind{norm}^q(\ell) \equiv
  S^q(\ell) / \left(S^2(\ell)\right)^{q/2}
\end{equation}
In particular, $S_\textind{norm}^4(\ell)$ is called the flatness, $F(\ell)$.

In phenomenological turbulence like \citet{k41a} or Iroshnikov-Kraichnan
\citep{iro63,krai65}, the structure functions of the velocity are power laws
of index $\zeta_q$ with $\zeta_q \propto q$, and this leads to normalized
structure functions that are constant as a function of the scale $\ell$.
When higher-order structure functions, in particular the flatness, depend on
$\ell$, it is a deviation from the above-mentioned classical phenomenologies
of turbulence and the field is intermittent.

\begin{figure}[tp]
  \centering
  \includegraphics[width=\linewidth,bb=29 0 415 284]{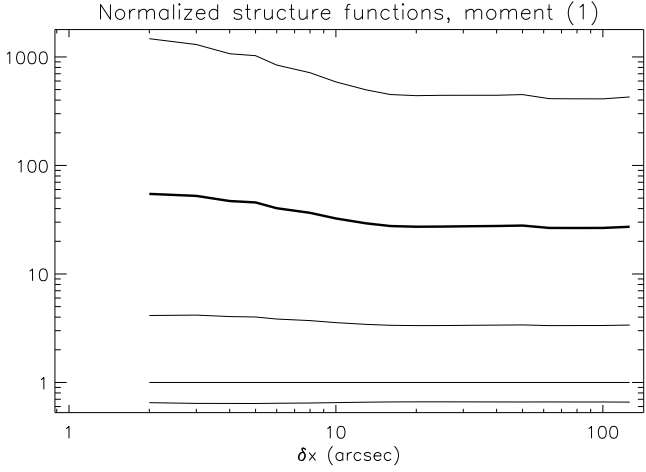}
  \caption{Normalized structure functions of index 1 to 5 (from bottom to
    top) for the field of moment~\moment{1} (\svii spectral radiance),
    averaged over 20 observations of the quiet Sun at disk center; the
    thick line is the flatness.}
  \label{fig:sumisfun}
\end{figure}

\paragraph{\svii spectral radiance.}
The normalized structure functions for the scalar field of moment 1 (\svii
spectral radiance) averaged over 20 observations of the quiet Sun at disk
center are shown in Fig.~\ref{fig:sumisfun}. They clearly rise when the
spatial scales gets smaller, especially when they get smaller than
$15\unit{Mm}$. This rise is characteristic of correlations in the field and
of intermittency, and can also be seen in the normalized structure functions
of the other radiance moments (of \lyeps and \svij). However, when studying
turbulent fields, the fields of which the intermittency is of primary
interest are the magnetic and velocity fields. Consequently we now examine
the latter.

\begin{figure}[tp]
  \centering
    \includegraphics[width=\linewidth,bb=29 0 415 284]{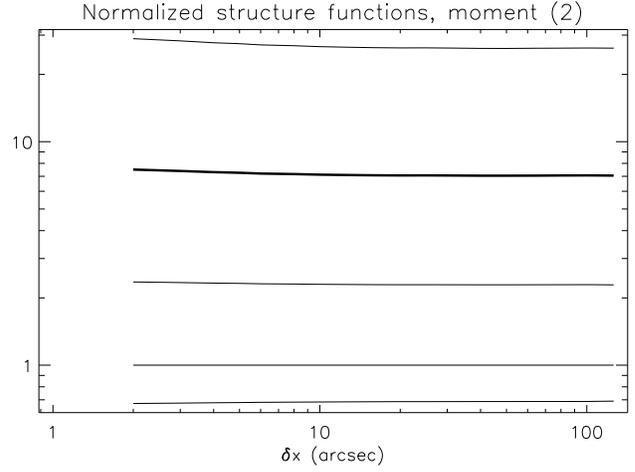} \\
    \includegraphics[width=\linewidth,bb=29 0 415 284]{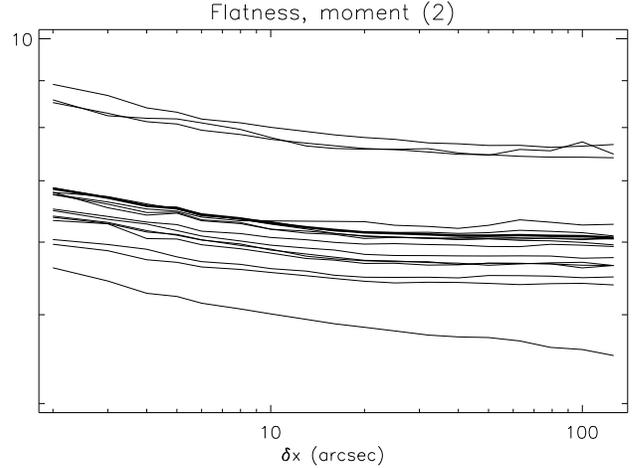} \\
   \caption{%
     Top: normalized structure functions of index 1 to 5 (bottom to top) for
     the field of moment~\moment{2} (\svii line shift), averaged over 16
     observations of the quiet Sun in the disk center; the thick line is the
     flatness. Bottom: flatness of the field of moment~\moment{2} at the
     dates used to compute the average (which is shown with the thick
     line).}
  \label{fig:sumvsfun}
\end{figure}

\paragraph{\svii line shift.}
When the field is a vector field $\vec a(\vec x)$, it is usually projected
longitudinally when computing the structure functions (in
Eq.~(\ref{eq:sfdef}), $\delta_{\vec \ell} a(\vec x)$ is replaced by
$\delta_{\vec \ell} a(\vec x) \cdot \vec\ell/\ell$). In our case, as well as
in magnetic field observation like \citet{abr02}, we have no choice but to
do this projection along the line-of-sight unitary vector
$\vec{e_\textrm{los}}$:
\begin{equation}
  \label{eq:sumfs}
  S^q(\ell)=
  \left\langle\left|
      (\vec{v}(\vec{x}+\vec{\ell}) - \vec{v}(\vec{x}))
      \cdot \vec{e}_\textrm{los}
    \right|^q
  \right\rangle_{\vec{x}, \,\|\vec{\ell}\|=\ell}
\end{equation}
The resulting average normalized structure functions are shown in top of
Fig.~\ref{fig:sumvsfun}. Because of the noise (which is not spatially
correlated) in the fields of moment~\moment{2} and perhaps because of other
effects discussed later, these structure functions are much flatter than in
the case of moment~\moment{1}, even if a small rise at the small scales can
still be seen. However, when plotting the flatnesses of the field of
moment~\moment{2} at all dates used to compute the average
(Fig.~\ref{fig:sumvsfun} bottom), one sees that the rise of the flatness is
systematic, and this gives a higher level of confidence in the rise observed
on average. It seems thus that we can observe intermittency in the velocity
field of the \svii line.

\section{Discussion}

Many difficulties arise when attempting to compute the statistics presented
here, mainly because of trade-offs between the need to get large fields at
high spatial resolution during a limited time (during which the Sun has not
changed much), and to:
\begin{itemize}
\item send all the required data to the ground with a given bandwidth (or to
  keep them temporarily in the limited SUMER memory)
\item have long enough exposure times so as to have enough signal.
\end{itemize}
Because of the first trade-off we need to compute the moments of the lines
on board and compress them by limiting their dynamics (see
Appendix~\ref{sec:comp}), and the second trade-off is at the origin of the
noise problem, which is particularly important for the line shift data
(moment 2). But with the current instruments there is no other alternative:
we think that the data set we used here is one of the most adapted to the
purpose.

Despite the difficulties, the tentative results presented here support the
presence in the transition region of small scales created by turbulence and
the intermittent nature of turbulent fields. To our knowledge this is the
first determination of the intermittency of the velocity field in an
ultraviolet line.  These small scales, at which dissipation is more
efficient, may play a role in the heating of the plasma of the solar upper
atmosphere, particularly of the corona (if confirmed by similar observations
in hotter lines).

Improved observations should be obtained with more sensitive spectrographs,
such as the ones from Solar~B and Solar Orbiter, which will allow us to make
progress by increasing spatial resolution while also increasing counting
statistics and thus allow, in particular, a better determination of the
velocity field.  However, there will still be compromises to be made on the
path to ``ideal'' observations which would have less noise, better spatial
and temporal resolution, and which would include lines emitted at hotter
tempratures further up in the corona, closer to where the heating is
believed to take place. One step further could be attained by using novel
ultraviolet imaging spectrograph designs, such as the MOSES spectro-imager
\citep{kankelborg01} or an Imaging UV Fourier Transform Spectrometer
\citep{millard04}.

\appendix

\newenvironment{code}%
  {\vspace*{-2\parskip}\begingroup\small\verbatim}%
  {\endverbatim\endgroup\vspace*{-2\parskip}}
\newcommand{\ccode}[1]{{\texttt{#1}}}

\def\near{\mathop{\rm near}\nolimits}
\def\byte{\mathop{\rm byte}\nolimits}

\section{Some details about the compression algorithm}
\label{sec:comp}

The need for sending to the ground the data of several full-Sun images (one
per moment) at a high resolution, in a limited time, and with a limited
bandwidth ($10.5\unit{kbps}$), forced us not only to compute the moments on
board (``compression'' method number 17 of the SUMER Operations Guide) but
also to limit the number of bits used for one moment in one pixel. The
principle of this second ``compression'' is simply to reduce the dynamics to
a 8-bit unsigned byte and to saturate high (and low, in the case of line
shifts) values.

To perform this, the function \ccode{near} is defined by the code
corresponding to:
\begin{equation}
  \near\,(x) = \byte\,(\min\,(x + 0.5, 255))
\end{equation}
where $\min\,(\bullet,\bullet)$ returns the smallest of its both arguments
and $\byte\,(\bullet)$ transforms its argument into a one-byte integer lower
approximate value.

If $M_1,\ldots,M_5$ are the 5 moments computed by the compression scheme
number 17 for one pixel of the image, then the values transmitted to the
ground are:
\begin{eqnarray}
  \label{eq:comp}
  M_1'&:=&\near\,(255\,M_1 / C_1) \\
  M_2'&:=&\near\,(\max\,(10\,M_2+128,0))\\
  M_3'&:=&\near\,(10\,M_3)\\
  M_4'&:=&\near\,(255\,M_4 / C_2) \\
  M_5'&:=&\near\,(255\,M_5 / C_3)
\end{eqnarray}
where $C_i$ is the \ccode{COMPAR}$i$ parameter given in the headers of the
FITS file.

The inverse operation is performed on the ground to recover approximations
of the original values of $M_1,\ldots,M_5$.

The on-board computation of the moments and the further compression of the
data were necessary to get the results presented in this paper from \sumer
data.  However, future observations of this kind should learn from the
limitations found here: even if they could still benefit from such a method,
there is a need for more signal in the original spectral data, and for a
compression scheme that allows a better dynamics (like a lossless
compression scheme applied on data with more than 8 bits).

\begin{acknowledgements}
  The authors acknowledge partial financial support from the PNST (Programme
  National Soleil--Terre) program of INSU (CNRS) and from European Union
  grant HPRN-CT-2001-00310 (TOSTISP network). They would like to thank Klaus
  Wilhelm and the anonymous referee for their valuable comments on this
  paper. The SUMER project is supported by DLR, CNES, NASA and the ESA
  PRODEX Programme (Swiss contribution). SoHO is a project of international
  cooperation between ESA and NASA.
\end{acknowledgements}

\bibliographystyle{aa}
\bibliography{solphys,needtoget}

\end{document}